\documentclass[%
floatfix,
reprint,
amsmath,amssymb,
pre,prr,aps,nofootinbib
]{revtex4-2}
\usepackage[english]{babel}

\usepackage{amsfonts,array}
\usepackage{bm}
\usepackage{dcolumn}
\usepackage[output-decimal-marker={.}]{siunitx} 

\usepackage[final]{graphicx} 
\usepackage{listings} 
\usepackage[usenames]{color} 
\usepackage{xcolor}
\usepackage{float}

\usepackage{hyperref}
\hypersetup{
	colorlinks=true,
	linkcolor={blue}, 
	filecolor=magenta,   
	citecolor={blue}, 
	urlcolor=blue, 
}

\usepackage{soul}

\date{\today}

\newcommand*{\Scale}[2][4]{\scalebox{#1}{$#2$}}%

\newcommand{\bx}{{\bf x}} 
\newcommand{\bk}{{\bf k}} 
\newcommand{\by}{{\bf y}}

\newcommand{\bn}{{\bf n}}
\newcommand{\bM}{{\bf m}}
\newcommand{\bz}{{\bf z}}
\newcommand{\bP}{{\bf P}}
\newcommand{\bp}{{\bf p}}

\begin{document}
	
	
	\title{Hydrodynamic description of proliferating active matter}
	
	\author{Nathan O. Silvano}
	\email{nathan@ifisc.uib-csic.es}
	\author{Emilio Hern\'andez-Garc\'\i a}
	\email{emilio@ifisc.uib-csic.es}
	\author{Crist\'obal L\'opez}
	\email{clopez@ifisc.uib-csic.es}
	\affiliation{Instituto de F{\'\i}sica Interdisciplinar
		y Sistemas Complejos (IFISC), CSIC-UIB, Campus Universitat de les
		Illes Balears, E-07122 Palma de Mallorca, Spain}
	
	\begin{abstract}
We develop a continuum theory for proliferating active matter starting from a microscopic stochastic model of self-propelled particles undergoing birth, death, and nonlocal competition. 
Beginning from the master equation, we derive mean-field evolution equations for the particle density and polarization fields and close the resulting hierarchy through a von Mises ansatz, providing a coupled hydrodynamic description applicable to a broad class of proliferating active systems. 
As an application, we study the recently introduced active Brownian bug model, in which the form of flocking emerges despite the absence of explicit alignment interactions. Linear stability analysis predicts both Turing and Hopf instabilities, whose analytical thresholds agree with numerical simulations. The continuum model reproduces the principal dynamical regimes of the underlying particle system, including homogeneous states, stationary periodic clusters, and coherently propagating flocking states. These results establish a general continuum framework for proliferating active matter and provide a physical interpretation of collective motion driven by the interplay between activity and population dynamics.

	\end{abstract}
	
	\maketitle
	

\section{Introduction}
	\label{Sec:intro}

The emergent properties of 
interacting self-propelled particles
 have become a major topic of study in recent decades, 
mainly driven by their many 
applications across biological, physical,
chemical, and artificial systems \cite{bechingerActiveParticlesComplex2016,royallColloidalHardSpheres2024,ramaswamyMechanicsStatisticsActive2010,tevrugtMetareviewSurveyActive2025,RevModPhys.85.1143}. 
Classical examples include animal flocking and schooling \cite{balleriniInteractionRulingAnimal2008,cavagnaBirdFlocksCondensed2014}, 
self-propelled Janus particles \cite{buttinoniDynamicalClusteringPhase2013}, 
and swarming bacteria \cite{sokolovConcentrationDependenceCollective2007}. In most of these 
theoretical frameworks, the number of particles 
is assumed to remain constant. However, this approximation
breaks down in many biological contexts, where individual
birth and death processes play an essential role. 
The interplay between individual proliferation and active 
motion is indeed fundamental to biological processes 
such as wound healing, tissue formation, 
and bacterial colony growth \cite{szaboPhaseTransitionCollective2006,alertPhysicalModelsCollective2020}. 
The emergent phenomena 
that may arise from the continuous creation
and removal of individuals,
coupled with self-propulsion,
include pattern formation, phase transitions,
morphogenesis, and collective migration, giving 
rise to the field of \textit{proliferating active matter} \cite{hallatschekProliferatingActiveMatter2023,almodovarLiquidhexaticsolidPhasesActive2022}.


In active matter, a particularly relevant phenomenon 
is flocking: the collective motion by which individuals move together
aligning their movements, as seen in bird flocks 
and fish schools. Many models exhibiting a flocking transition,
starting from the  Vicsek model \cite{vicsekmodel1995},
 incorporate an explicit alignment interaction \cite{martinTransitionCollectiveMotion2025,romanczukActiveBrownianParticles2012,baconnierSelfaligningPolarActive2025}. 
However, flocking behavior in the absence 
of explicit alignment interactions has also been 
reported \cite{romanczukActiveBrownianParticles2012,capriniSpontaneousVelocityAlignment2020, salahshourAllocentricFlocking2025, capriniFlockingAlignmentInteractions2023, dasFlockingTurningAway2024}, showing that 
such interactions are not strictly 
necessary for the emergence of flocking. 
Here we draw attention to a recent result 
\cite{v3gm-vhry} in which flocking is induced 
by proliferation dynamics, with no explicit alignment 
interaction. The individual-based model (IBM) is built 
upon the neighborhood-dependent 
birth-rate
population model of Ref.~\cite{ehgclopezpre}, 
coupled to active motion. This result 
highlights the role of proliferation, when coupled to self-propulsion,
in the emergence of novel collective phenomena, 
and underlines the need for 
theoretical approaches to study 
proliferating active matter.

In this work, we progress along this line 
by introducing a generalized model 
for self-propelling active particles with reproduction, 
death, and spatially nonlocal 
competition for resources. We derive a hydrodynamic description
in terms of mean-field equations under a closure ansatz leading to a description in terms of the 
particle density and polarization field. This framework
is then applied to the specific proliferating active 
matter model of Ref.~\cite{v3gm-vhry}, providing 
a  continuum approximation to it. We show that 
the continuum model accurately captures most of 
the different dynamical regimes of
the underlying discrete particle dynamics, and we also highlight that
 the continuum description is used to
interpret the physical mechanisms responsible for each phase.

The outline of the paper is the following.
In Sec.~\ref{Sec:Model}, we propose a generalized proliferating active 
particle model and derive its continuum description. Most of the technical details are provided in the Appendixes. In Sec.~\ref{Sec:ABB}, 
we apply this to the model in Ref.~\cite{v3gm-vhry} and study (analytically and numerically) our continuum  description
in one and two dimensions.
In Sec.~\ref{Sec:summary}, we give a summary and conclusions of this work. 	

\section{Model}\label{Sec:Model}	
	
We present here our general proliferating particle model and then obtain 
a continuum description.	
	
\subsection{Particle dynamics}

\begin{figure}[htb!]
	\includegraphics[width=\columnwidth]{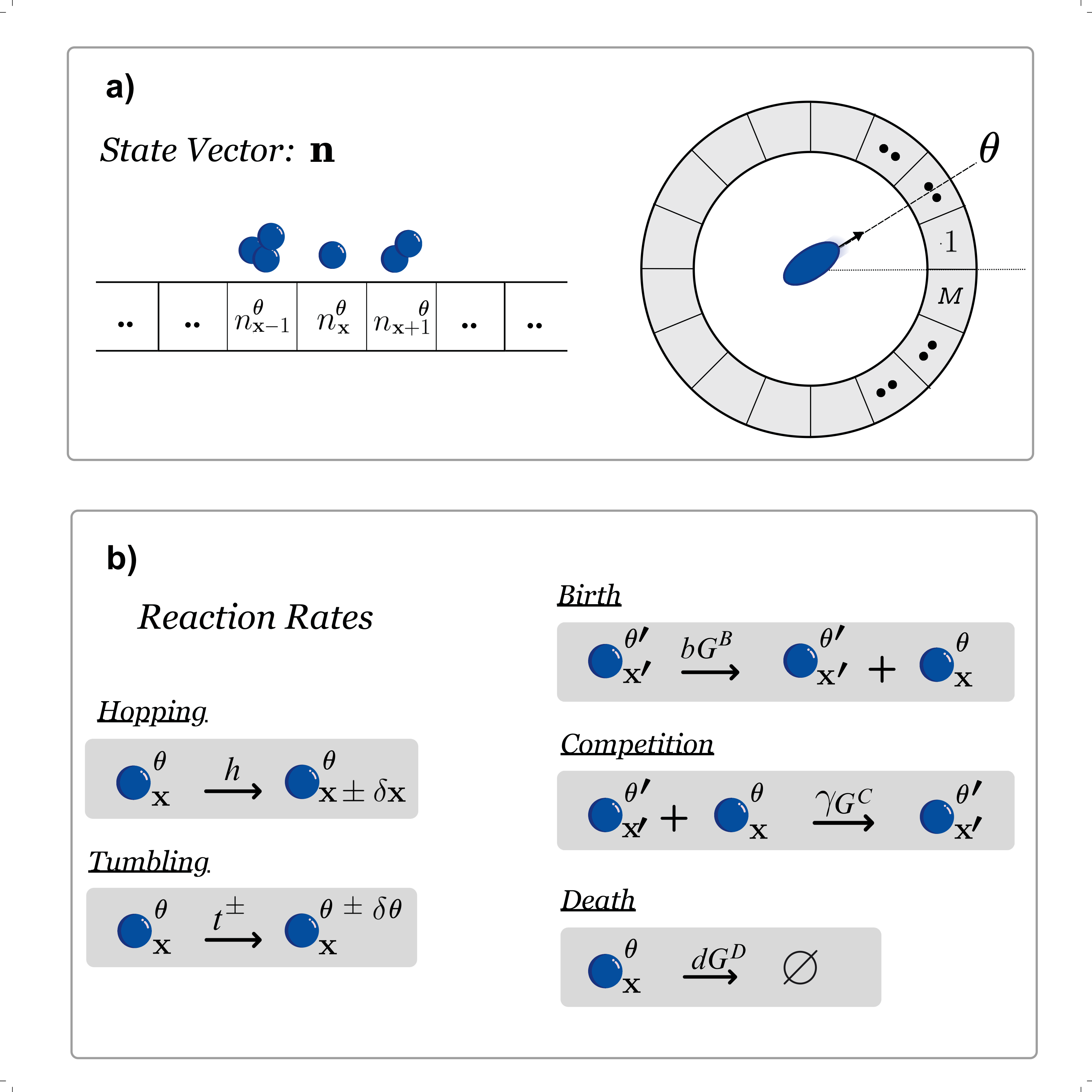}
	\caption{Microscopic model representation. \textbf{a)} Schematics of the state vector $\bn$. Left: Spatial lattice with occupation numbers $n_\bx^\theta$. Right: Discretized orientation space for a single particle in two dimensions. \textbf{b)} Individual-based model reaction rates, including hopping and tumbling dynamics, as well as birth, competition, and death processes.}
	\label{fig:micro_model}
\end{figure}

We consider a lattice-based microscopic model
for active particles. For simplicity we restrict
the general discussion to two 
spatial dimensions and a square lattice. The state 
of each particle is characterized by a pair $\bx,\theta$, 
where ${\bf x} = (x,y)$ is the discrete spatial 
position on the lattice and $\theta$ is a
discrete orientation variable 
(Fig.~\ref{fig:micro_model}a) with $M$ possible values. 
The orientation will determine, in the 
continuum limit to be defined below, 
the direction of motion of the particle, 
giving rise to active motion. We allow 
multiple occupation of the lattice sites. 
Thus, the microscopic state of the full 
system of $N$ particles is specified by the array 
\begin{align*}
\bn \equiv \{n^\theta_{\bf x}\}_{({\bf x},\theta)} \ , 
\end{align*}
where each occupation number $n_\bx^\theta \in \mathbb{N}$ gives the 
number of particles at site $\bf x$ with
orientation $\theta$. 
Note that if we consider, as we will do in some 
numerical studies, 
only one spatial dimension, one may restrict to 
$M=2$ internal states  (left- or right-moving). 
However, throughout this work we allow 
for a general number $M$ of internal states even in $1D$, 
corresponding to a free planar orientation degree of freedom.
In one dimension this describes self-propelled rotors that
can move only left or right, but in function of a richer instantaneous orientation \cite{demaerelActiveProcessesOne2018}.

Each particle undergoes a minimal set of stochastic
processes, summarized in Fig.\,\ref{fig:micro_model}b. First, 
in a reproduction event, a particle in state $(\bx',\theta')$ 
can give birth to a new one in state $(\bx,\theta)$ with a 
reproduction rate (\textit{i.e.} probability per unit of time) per particle $b G^B_{\bx' \bx\theta'\theta}(\bn)$. We have made explicit that the birth rate $G^B$ can depend on the state $\bn$ of the whole system, although we will allow later a dependence only on the neighborhood of the reproducing particle at $(\bx',\theta')$. 
Second, the particle can die with rate $d G^D_{\bx\theta}(\bn)$, which in general can depend on the occupation numbers of the whole lattice, although only sufficiently close neighbors are expected to give significant influence. 
Finally, a particle in state $(\bx',\theta')$ can interact competitively with other particles in state $(\bx,\theta)$ (later on we will restrict to locations $\bx'$ sufficiently close to $\bx$) which would lead to the death of the particle at $(\bx,\theta)$. The rate of this process for each pair of particles would be $\gamma G^C_{\bx' \bx\theta'\theta}(\bn)$. Later, in the explicit applications, normalization for the rates $G^{B,D,C}$ will be chosen in such a way that the parameters $b,d,\gamma$ characterize the main inverse time scales. 

Concerning motility, each particle can hop to neighboring lattice sites at rates $h^R$, $h^L$, $h^U$, $h^D$ (for jumps to right, left, up and down, respectively). An asymmetry between these hopping rates (\textit{e.g}. $h^R \neq h^L$) introduces a finite 
propulsion velocity that will be linked to the orientation variable $M$. Reorientation dynamics (tumbling) is modeled by transitions between neighboring orientation states $M\rightarrow M\pm 1$, occurring at rates $t^+$ and $t^-$ per particle. If the tumbling rates are taken to be symmetric in 
orientation space ($t^+=t^-$), the standard active Brownian particle (ABP) limit
\cite{digregorio} is obtained in a continuous limit (see Apps.~\ref{sec:app_masterEq} and \ref{sec:continuum_MF}).

The microscopic dynamics is governed by a master equation for the probability $P(\bn,t)$ of finding the system in state $\bn$ at time $t$,
\begin{align}
	\frac{\partial P(\bn,t)}{\partial t} 
	= \sum_{\bn'} \Big[ \Omega(\bn' \rightarrow \bn)\,P(\bn',t) 
	- \Omega(\bn \rightarrow \bn')\,P(\bn,t) \Big],
\end{align}
where the total transition rate $\Omega$ gets contributions from the different microscopic processes,
\begin{align}
	\Omega = W_B + W_D + W_C + W_H + W_T \, .
\end{align}

Taking into account the above definitions for the rates per particle and per pair, the only non-vanishing transition rates associated to demographic process ($W_B$, $W_D$ and $W_C$ for birth, death and competition, respectively) are given by
\begin{align}
	&W_B(\bn \rightarrow \bn+ \bm{e}_{\bf x}^{\theta}) 
	= b\, \sum_{\bx'\theta'} G^B_{\bx' \bx\theta'\theta}(\bn) \, n_{\bx'}^{\theta'} ,
	\label{eq:birth_transition_rate}
	\\
	&W_D(\bn \rightarrow \bn - \bm{e}_{\bf x}^{\theta}) 
	= d\,G^D_{\bx\theta}(\bn)\, n_{\bx}^{\theta} ,
	\label{eq:death_transition_rate}
	\\
	&W_C(\bn \rightarrow \bn - \bm{e}_{\bf x}^{\theta}) 
	= \gamma\, \sum_{\bx'\theta'} G^C_{\bx' \bx\theta'\theta}(\bn) \,
	n_{\bf x}^\theta n_{\bf x'}^{\theta'} ,
	\label{eq:comp_transition_rate}
\end{align} 
where $\bm{e}_{\bf x}^{\theta}$ is an array of the same size as $\bn$, with all components equal to zero except the component $(\bx, \theta)$, which has value 1. Thus, $\bn + \bm{e}_{\bf x}^{\theta}$ is the state $\bn$ of the system modified by one additional particle at location $(\bx, \theta)$. The sums in $W_B$ and $W_C$ implement the fact that the birth by reproduction and the death by competition of one particle at state $(\bx,\theta)$ can arise from interactions with many particles at different states $(\bx',\theta')$. 

The hopping transition rate $W_H$ gets contributions from the jump rates per particle from location $\bx$ (and particle orientation $\theta$) towards the four grid directions, $h_{\bx\theta}^R$, $h_{\bx\theta}^L$, $h_{\bx\theta}^U$ and $h_{\bx\theta}^D$. The superindices indicate the direction of the jump, towards right, left, up and down, respectively. We assume that the size of the jumps is just one grid size. To describe some self-propulsion dynamics, it is essential that these rates are coupled to the orientational degree of freedom. This coupling will be specified in Appendix \ref{sec:continuum_MF}, Eqs. (\ref{app:startjumprates})-
(\ref{app:endjumprates}), and commented in the next Section.

The transition rates for  tumbling $W_T$, which we assume to involve only transitions from $M$ to $M\pm 1$,  
are $W_T=W_T^+ + W_T^-$ which in terms of the per-particle rates from orientation $\theta$ and location $\bx$, $t_{\bx\theta}^+$ and $t_{\bx\theta}^-$, read  
\begin{align}
	W_T^+(\bn \rightarrow \bn - \bm{e}_{\bf x}^{\theta} + \bm{e}_{\bf x}^{\theta+1}) 
	= t_{\bx\theta}^{+}\, n_{\bf x}^\theta ,
	\\
	W_T^-(\bn \rightarrow \bn- \bm{e}_{\bf x}^{\theta} + \bm{e}_{\bf x}^{\theta-1}) 
	= t_{\bx\theta}^{-}\, n_{\bf x}^\theta .
	\label{eq:state_dep_tumbling_transition_rate}
\end{align}

Our goal is to obtain a continuum description that, under reasonable assumptions and approximations, properly describes this IBM dynamics. The first step, after having defined the master equation, is to take the average-density description. The mean number of particles at site ${\bf x}$ with orientation $\theta$ is defined as $\langle n_{\bf x}^\theta \rangle 
= \sum_{\bn} n_{\bf x}^\theta \, P(\bn,t)$, so that its temporal evolution is (See App.~\ref{sec:app_masterEq} for a explicit derivation)
\begin{widetext}
\begin{align}
	\frac{d\langle n_{\bf y}^\beta\rangle}{dt}	 
=& b\sum_{{\bf x'} \theta'} \left\langle G^B_{ \bx'\by\theta'\beta}(\bn) n_{\bx'}^{\theta'} \right\rangle 
- d\left\langle G^D_{\by\beta}(\bn) n_\by^{\beta} \right\rangle - \gamma\sum_{\bx'\theta'} \left\langle G^C_{\bx'\by\theta'\beta}(\bn) n_\by^\beta n_{\bx'}^{\theta'} \right\rangle 
	\nonumber\\
&+ \sum_{\alpha\in\{R,L,U,D\}} \left[ h^{\alpha}_{\mathbf{y}-\boldsymbol{\varepsilon}_\alpha,\beta} \langle n_{\mathbf{y}-\boldsymbol{\varepsilon}_\alpha}^\beta \rangle - h^{\alpha}_{\mathbf{y},\beta} \langle n_{\mathbf{y}}^\beta \rangle \right]   
	+
	 t^+_{\by \beta-1}\langle n_{\bf y}^{\beta-1}\rangle
	 +
	 t^-_{\by \beta+1}\langle n_{\bf y}^{\beta+1}\rangle
	 -
	 (t^+_{\by \beta} + t^-_{\by \beta})\langle n_{\bf y}^\beta\rangle.
\label{eq:formeanN}
\end{align}
\end{widetext}
$\boldsymbol{\varepsilon}_\alpha$, with $\alpha\in\{R,L,U,D\}$ are the lattice unit vectors in the right, left, up and down directions. In addition to the assumption of single-step jumps for the spatial hopping and angular tumbling rates, we have also assumed that they are independent of the particular system occupation numbers, $\bn$, so that they are not affected by the averages. 
	
	\subsection{Mean-Field and Continuum Equations}
	
Since Eq. (\ref{eq:formeanN}) couples the mean occupation to higher order moments, some approximation is needed to obtain from it a useful expression. The first approximation we introduce is a mean-field approximation in the nonlinear terms, so that averages are factorized:  $\langle n_\by^\beta n_{\bx'}^{\theta'} \rangle \approx \langle n_\by^\beta\rangle \langle n_{\bx'}^{\theta'} \rangle$, $\langle G(\bn) n_\by^\beta \rangle \approx \langle G(\bn)\rangle \langle n_\by^\beta \rangle$, etc.  

Next, we introduce a continuum description by replacing the average occupation numbers by an average density $\rho(\bx,\theta,t)=\langle n_\bx^\theta \rangle /(\Delta\theta \Delta x^d)$, where $\Delta \theta$ is the angular separation between consecutive allowed values of the orientation angle, and $\Delta x$ is the lattice step value ($d$ is the lattice dimension, here mostly taking the value $d=2$), and taking the limit $\Delta x, ~\Delta\theta \rightarrow 0$. Essential to the modeling of active self-propulsion we want to implement, is the coupling between spatial hopping and angular orientation. As shown in Appendix \ref{sec:continuum_MF}, a standard model of active-matter motion \cite{digregorio}, is recovered by introducing an  asymmetric component, coupled to the particle orientation, distinguishing between left and right and between up and down jump rates on the lattice. This asymmetry has been taken to be proportional to the projection of the orientation direction of the particle on the corresponding lattice direction. For example, the left and right jump rates would have the form $h_{\bx\theta}^{R,L}=A\pm B v_0 \cos\theta$. {We stress that in the square lattice there are always four possible jump directions, with the corresponding four jump rates, but we can have the angle $\theta$ discretized in an arbitrary number $M \geq 4$ of values, being the continuum limit attained for $M\rightarrow\infty$}.The strength of the asymmetry determines the parameter $v_0$, which will become the self-propulsion velocity. Additional details for the continuum limit of this and of the remaining terms in the equation of motion are in Appendix \ref{sec:continuum_MF}, where we obtain all the terms describing the dynamical evolution of the spatial and angular mean density in this continuum limit: 
	\begin{widetext}
	\begin{align}
		\frac{\partial \rho({\bf x},\theta,t)}{\partial t} =&  \int d{\bf y} d\beta\; \rho({\bf y},\beta,t) \bigg\{b\,G^B(\by,\bx,\beta,\theta)  - \gamma \rho({\bf x},\theta,t)   G^C(\by,\bx,\beta,\theta)  \bigg\}
		\nonumber	\\
		&- d~ G^D(\bx,\theta)\rho({\bf x},\theta,t)
-v_0 \boldsymbol{\nabla}\cdot(\hat{\mathbf{n}}(\theta)  \rho({\bf x},\theta,t)) 
+ D_T\boldsymbol\nabla^2\rho({\bf x},\theta,t) 
+ D_R \partial^2_\theta \rho({\bf x},\theta,t) \ ,
		\label{eq:final_result}
	\end{align}
	\end{widetext}

{where $\hat{\bn}(\theta)=(\cos\theta,\sin\theta)$ is the heading unit vector.} Eq.\,\eqref{eq:final_result} is a quite general equation for the temporal evolution of the average spatial and orientational density of active Brownian particles.  In this equation, the kernels $G^D$ and $G^C$ are simply the corresponding rates in (\ref{eq:death_transition_rate}) and (\ref{eq:comp_transition_rate}) under a change of notation. The birth rate needs a renormalization: {$G^B(\by,\bx,\beta,\theta)=G^B_{\by\bx\beta\theta}/\left(\Delta\theta(\Delta x)^d\right)$}. For notational convenience, in these three rates of the continuum equation we do not indicate any dependence on the detailed occupation state $\bn$ of the system, although there is no problem (after the mean-field approximation has been done) for it to be included. In addition, as shown in the Appendix \ref{sec:app_masterEq}, a negative linear dependence of $G^B$ on $\bn$, as well as a positive linear dependence of $G^D$ on $\bn$, has a competitive character and can be simply absorbed in a redefinition of $G^C$. 
As commented above, the advection term containing $v_0$ represents self-propulsion, and arises from the asymmetric left-right and up-down component, coupled to the particle orientation, of the  jump rates on the lattice. The symmetric component of the jump rates gives rise to the translational diffusion term containing $D_T$. Tumbling transitions are assumed to be symmetric for $M\rightarrow M\pm 1$ and give rise to the term containing the rotational diffusion coefficient $D_R$. $D_T$ and $D_R$ have been assumed to be spatially and angularly independent, as well as independent on the occupations $\bn$. Additional details are in Appendix \ref{sec:continuum_MF}. 


	\subsection{Angular moments and closure approximation}

	A standard way  of working with the orientational dependence of the 
	density $\rho({\bf x},\theta,t)$ is to introduce the angular Fourier representation \cite{delfauActiveClusterCrystals2017,bertinHydrodynamicEquationsSelfpropelled2009,gautryClosuresMomentExpansion2026,RevModPhys.85.1143},
\begin{equation}
	\rho_n({\bf x},t) = \int_0^{2\pi} d\theta\, \rho({\bf x},\theta,t) e^{in\theta} ,\quad \text{for} \;\;n=0, \pm 1, \pm 2, ...  \label{eq:rho_n}
\end{equation}
as well as the inverse representation,
\begin{equation}
	\rho({\bf x},\theta,t) = {\frac{1}{2\pi}}\sum_{m \in \mathbb{Z}} \rho_m({\bf x},t) e^{-i m \theta}. \label{eq:rho_n_inverse}
\end{equation}

With these definitions we can rewrite Eq.\,\eqref{eq:final_result} as a coupled set of equations for the angular modes (or moments) $\rho_n(\bx,\theta)$ (see Appendix C)::
	\begin{align}
		\partial_t \rho_n = \Delta_n^B +\Delta_n^D + \Delta_n^C +D_T\nabla^2 \rho_n - n^2D_R\rho_n \nonumber
		\\- \frac{v_0}{2}\bigg\lbrace (\partial_x -i\partial_y)\rho_{n+1} +(\partial_x+i\partial_y)\rho_{n-1}\bigg\rbrace \label{eq:rhon_modes}.
	\end{align}
We have used $\bx=(x,y)$. $\Delta_n^B$, $\Delta_n^D$, and $\Delta_n^C$ are the angular Fourier transforms of the birth, death and competition terms, respectively, in Eq.\,\eqref{eq:final_result}. To further evaluate them, we now add some simplifying assumptions. 
First, we assume that the birth and competition kernels can be factorized into a spatial and an orientational part, $G^{B,C}(\by,\bx,\beta,\theta) = S^{B,C}(\by,\bx) \Phi^{B,C}(\beta,\theta)$. Second, as already imposed for $D_T$ and $D_R$, we assume spatial and orientational homogeneity and isotropy for the demographic terms. {This implies that the death rate becomes a constant, and the magnitude of $G^D(\bx,\theta)$ can be absorbed in the parameter $d$}. Also, the spatial and angular kernels for birth and competition would depend only on the modulus of differences of positions and angles, respectively, $S^{B,C}(\by,\bx)=S^{B,C}(|\by-\bx|)$, $\Phi^{B,C}(\beta,\theta)=\Phi^{B,C}(|\beta-\theta|)$. Thus, the corresponding contributions to Eq.\,\eqref{eq:rhon_modes} read  (See Appendix  \ref{app:momentum_exp}): 
%

\begin{align}
&\Delta^B_n(\bx,t) = b~\Phi^B_n~ \left( S^B * \rho_n \right)(\bx,t) \ , \label{eq:DeltaB}
\\ &\Delta^D_n(\bx,t) = - d~\rho_n(\bx,t)    \,,\label{eq:DeltaD}
\\ &\Delta^C_n(\bx,t) = -\frac{\gamma}{2\pi} \sum_m \rho_{n-m}(\bx,t) \left( S^C*\rho_m \right)(\bx,t)~ \Phi^C_m \ ,
\label{eq:DeltaC}
\end{align}
where we have introduced the Fourier transform of the angular kernels,  $\Phi^{B,C}_n = \int d\theta~ \Phi^{B,C}(\theta) e^{in\theta}$, and the definition of spatial convolution, $\left(S^{B,C}*\rho_n\right)(\bx)=\int d\by S^{B,C}(\bx-\by)\rho_n(\by)$.

The angular Fourier moments $\rho_n(\bx,t)$ defined in Eq.\,\eqref{eq:rho_n}
	have a direct physical meaning for the first two orders $n=0,\pm 1$. The zeroth
	moment is simply the mean spatial density of the particles,
	\begin{align}
		\rho_0(\bx,t) = \int_0^{2\pi} \rho(\bx,\theta,t)\,d\theta = \rho(\bx,t).
	\end{align}
	The first moment encodes the average orientational bias of the population at each point. To see this we define the \emph{polarization vector field}
in terms of $\hat{\bn}(\theta)=(\cos\theta,\sin\theta)$,
\begin{align}
	\bP(\bx,t) = \int_0^{2 \pi} \hat{\bn}(\theta) \rho(\bx,\theta,t) d\theta,
\end{align}
whose components are explicitly,
\begin{align}
	&	P_x(\bx,t) = \int_0^{2 \pi} \cos \theta \rho(\bx,\theta,t) d\theta\ , 
	\\
	&	P_y(\bx,t) = \int_0^{2 \pi} \sin \theta \rho(\bx,\theta,t) d\theta\ .
\end{align}
	$\bP$ measures the local net heading of the population, weighted by density:
	$\bP=0$ for an isotropic (unpolarized) angular distribution, and
	$|\bP|\to\rho$ as the population becomes fully aligned along a single
	direction. In a complex representation, $\bP$ is directly related to  the first angular Fourier mode,
	\begin{align}
		\rho_1(\bx,t) = \int_0^{2\pi} e^{i\theta}\rho(\bx,\theta,t)\,d\theta = P_x(\bx,t)+iP_y(\bx,t).
	\end{align}
$\rho_{-1}$ is just the complex conjugate of $\rho_1$, carrying thus the same information. 

$\rho_0$ and $\rho_{\pm 1}$ alone do not close the hierarchy in 
	Eq.\,\eqref{eq:rhon_modes}, the $n=\pm 1$ equation involves $\rho_{\pm 2}$ respectively, the
	$n=\pm 2$ equation involves $\rho_{\pm 3}$, and so on. Due to this fact, some assumption about the
	higher moments is needed to obtain a finite self-contained set of equations. We aim at a description in terms of $\rho$ and $\bP$. The simplest choice is a first-order truncation,
	discarding all $\rho_n$ with $|n|>1$, which amounts to approximating the
	angular distribution by its first harmonics,
	\begin{align}
	\rho(\bx,\theta,t)\approx \frac{\rho}{2\pi} + \frac{1}{\pi}\,\bP\cdot\hat{\bn}(\theta).	
\label{eq:1stTruncation}
	\end{align}
This is a standard route in active-matter kinetic theories
\cite{delfauActiveClusterCrystals2017,bertinHydrodynamicEquationsSelfpropelled2009,gautryClosuresMomentExpansion2026}. Nonetheless, the approximation is only a genuine probability density (non negative) for weak polar order, $|\bP|/\rho \le 1/2$, and breaks down precisely in the
strongly polarized regimes where flocking behavior is most relevant. We instead close the hierarchy with a von Mises ansatz for the angular
shape of the density. Namely, we assume that the expected  angular distribution at each time and spatial point follows a von Mises distribution \cite{kurstenCriticalAssessmentMises2017,lamSelfpropelledHardDisks2015,rodriguez-lujanRegularizedMultivariateMises2015,kumarFiniteTimeOrientationalRelaxation2026}, $f_{VM}(\theta;\phi,\kappa)$: 
	\begin{align}
		\rho(\bx,\theta,t) & \equiv \rho(\bx,t)\, f_{VM}\big(\theta;\phi(\bx,t),\kappa(\bx,t)\big)
		\nonumber\\	&= \frac{\rho(\bx,t)}{2\pi I_0(\kappa(\bx,t))}\, \exp\Big[\kappa(\bx,t)\cos\big(\theta-\phi(\bx,t)\big)\Big]\ .
		\label{eq:vonmises_ansatz}
	\end{align}

The von Mises distribution is a unimodal probability density (with support in $\theta \in [\phi-\pi,\phi+\pi]$), appropriate for angular variables, characterized by two parameters, $\phi$, the mean value of the angle, and $\kappa>0$ (usually called the \emph{concentration})  giving the inverse width of the distribution. Here, these two parameters are dependent on space and time, $\phi=\phi(\bx,t)$, $\kappa =\kappa(\bx,t)$. $I_n$ is the modified Bessel function of the first kind of integer order $n$. This is the maximum-entropy distribution on the circle compatible with a prescribed mean direction and angular variance \cite{biswasEfficientSamplingCircular2024}, and thus it can be considered an analogue of the Gaussian distribution for angular variables. Like for the first-order truncation, Eq.\,\eqref{eq:1stTruncation}, we will see that it becomes fully  
specified by $\rho$ and $\bP$ alone, but unlike the first-order truncation it remains always normalized and non-negative, making it a natural closure to carry through the rest of this work. We expect it to be a good approximation to the true $\rho(\bx,\theta,t)$ as far as the angular dependency has a single maximum \cite{kurstenCriticalAssessmentMises2017}. 
	
Substituting Eq.\,\eqref{eq:vonmises_ansatz} into the definition
Eq.\,\eqref{eq:rho_n} fixes every angular Fourier moment in terms of just
three scalar fields, $\rho(\bx,t)$, $\phi(\bx,t)$ and $\kappa(\bx,t)$: 
	\begin{equation}
		\rho_n(\bx,t) = \rho(\bx,t)\,\frac{I_n(\kappa(\bx,t))}{I_0(\kappa(\bx,t))}\, e^{in\phi(\bx,t)}.
		\label{eq:vonmises_moments}
	\end{equation}
Evaluating Eq.\,\eqref{eq:vonmises_moments} at $n=1$ and comparing with
$\rho_1=P_x+iP_y$ gives
	\begin{align}
		&P_x = \rho\,\mathcal A(\kappa)\cos\phi,
		\qquad P_y = \rho\,\mathcal A(\kappa)\sin\phi,
		\label{eq:P_kappa}	\\
		& \textrm{with} \ \ \ \mathcal A(\kappa)\equiv\frac{I_1(\kappa)}{I_0(\kappa)}.
\label{eq:Akappa}
	\end{align}
$\mathcal A(\kappa)$ is a monotonically increasing function of $\kappa$. It ranges from $\mathcal A(\kappa)=0$ for $\kappa=0$, in which case the von Mises distribution becomes a uniform angular distribution representing complete orientational disorder, to $\mathcal A(\kappa)\to 1$ as $\kappa\to\infty$, leading to a delta distribution peaked at $\theta=\phi$, meaning a perfect angular alignment. From \eqref{eq:P_kappa} we find
	\begin{align}
		\frac{|\bP(\bx,t)|}{\rho(\bx,t)} &= \mathcal A\big(\kappa(\bx,t)\big) \ ,  		
\label{eq:order_param}
\\  \phi &=\arg(P_x+iP_y) \ .
\label{eq:local_heading}
	\end{align}
The first equation reveals that the strength of the local polar order
is controlled entirely by $\kappa$, while the second indicates that $\phi$ gives the local mean polarization direction. 

For small $\kappa \ll 1$ (\textit{i.e.}, in the places where the angular distribution is close to uniform), $I_n(\kappa)\simeq(\kappa/2)^n/n!$, so that $\mathcal A(\kappa)\simeq \kappa/2$ and Eq.~\eqref{eq:vonmises_ansatz} reduces to
\begin{align}
	\rho(\bx,\theta,t) \approx \frac{\rho(\bx,t)}{2\pi}\Big[1+
\kappa(\bx,t)\cos\big(\theta-\phi(\bx,t)\big)\Big] \ ,
\end{align}
thus recovering the dipolar (truncated Fourier)
approximation (\ref{eq:1stTruncation}) with the proper approximation at small $\kappa$ to $\bP$ in (\ref{eq:P_kappa}), $\bP \approx (\rho \kappa/2) (\cos \phi ,\sin \phi)$. The von Mises closure is therefore a genuine generalization of
the first-order truncation: it agrees with it for weak polar order
($\kappa\ll1$) but, unlike the linear form, remains a suitable probability
density, positive and normalized, even for strongly polarized
regimes in which the angular distribution is very narrow. {An assessment of the accuracy of the von Mises closure (and of other approaches) to represent the angular distribution of self-propelled particles can be found in Ref.\cite{kurstenCriticalAssessmentMises2017}}.

To obtain the equations of motion of the densities describing the system we return to Eq.\,\eqref{eq:rhon_modes}. The $n=0$ equation of the hierarchy 
couples only to $\rho_0,\rho_{\pm 1}$ and can therefore be written exactly  irrespective of any closure approximation. It results in a continuity equation for $\rho$:
\begin{align}
	&\partial_t \rho = D_T\boldsymbol\nabla^2\rho - v_0\boldsymbol\nabla\cdot\bP + \Delta_0^B+\Delta_0^C - d\rho \;;
	\label{eq:rho_final}
	\\
&\Delta^B_0 =  b\Phi^B_0 \left( S^B*\rho_0 \right)(\bx,t) 
	\\
	&\Delta_0^C = - \frac{\gamma}{2\pi}  \sum_m \rho_{-m}(\bx,t) \left( S^C*\rho_m \right)(\bx,t) \Phi_m
\end{align}
Details of the derivation can be found in Appendix \ref{app:momentum_exp}. 

The $n=1$ equation, however, involves $\rho_2$. Appendix \ref{app:momentum_exp} shows than the von Mises ansatz gives an expression for it in terms of the fields $\rho_0(\bx,t)$ and $\rho_1(\bx,t)$. Using the equations for $n=\pm 1$, and the closure for $\rho_{\pm 2}$, the resulting equation of motion for the polarization vector $\bP(\bx,t)$ is (see Appendix \ref{app:momentum_exp}):

\begin{align}
&	\partial_t \bP =  D_T\nabla^2\bP - D_R\bP + \Delta_1^B + \Delta_1^C - d~\bP \nonumber \\
&	\qquad\qquad -\frac{v_0}{2} \boldsymbol\nabla\rho
	+ v_0 \boldsymbol\nabla\cdot  { \,\bf Q}  \ ,
	\label{eq:P_final}
		\\
	&\Delta^B_1 =  b\Phi^B_1 \left( S^B*\rho_1 \right)(\bx,t) 
	\\
	&\Delta_1^C = - \frac{\gamma}{2\pi}  \sum_m \rho_{1-m}(\bx,t) \left( S^C*\rho_m \right)(\bx,t) \Phi_m
\end{align}
which involves a quadrupolar, nematic-like, contribution in terms of a symmetric, traceless nematic tensor
\begin{align}
	Q_{ij}[\rho,\bP] = \rho(\bx,t)\mathcal B(\kappa(\bx,t))
	\Big(\frac{P_i(\bx,t)P_j(\bx,t)}{|\bP|^2} -\frac{1}{2}\delta_{ij}\Big) \ , 
	\label{eq:T_tensor}
\end{align}
with $i,j=x,y$. The function $\mathcal B(\kappa)$ is defined as
\begin{align}
	\mathcal B(\kappa)\equiv \frac{ I_2(\kappa)}{I_0(\kappa)}  \ .
	\label{eq:Bkappa}
\end{align}
%


Then, the procedure to solve the coupled set of equations (\ref{eq:rho_final}) and (\ref{eq:P_final}) involves the following steps: First, from initial density and polarization fields, $\rho(\bx,t)$ and $\bP(\bx,t)$, compute the fields $\kappa(\bx,t)$ via Eq.\,\eqref{eq:order_param} and $\phi(\bx,t)$ via Eq.\,\eqref{eq:local_heading}. The calculation of $\kappa$ can be done by numerical inversion since, according to its definition (\ref{eq:Akappa}), $\mathcal A$ is a monotonic function. Second, substitute $\kappa(\bx,t)$, $\rho(\bx,t)$ and $\bP(\bx,t)$ in (\ref{eq:Bkappa}) and (\ref{eq:T_tensor}) to obtain the tensor ${\bf Q}[\rho,\bP]$ at each point, and finally use the evolution equations (\ref{eq:rho_final}) and (\ref{eq:P_final}) to advance $\rho$ and $\bP$ one time step. 

\section{The Active Brownian Bug Model}\label{Sec:ABB}
	
To demonstrate the insights provided by our 
generalized particle model and its 
continuum description, we apply them to a 
recently introduced model, the so-called Active Brownian
 Bug (ABB) model \cite{v3gm-vhry}. 
This model is the active version
of the Neighborhood-dependent (ND) Brownian particle
 model from \cite{ehgclopezpre}, 
 extended to incorporate self-propelled motion. 
 Briefly, the model (in 2D) considers
an ensemble of $N(t)$ particles 
(the number changes with time) whose positions ${\bf r}_i =(x_i,y_i)$ evolve in continuous space as \cite{digregorio}:
\begin{align}
		&\frac{d{\bf r}_i}{dt} = v_0 \hat{\bf n}(\theta) + \xi_T^i, \label{eq:selfprop1}
		\\
		&\frac{d \theta_i}{dt} = \xi_r^i,
\label{eq:selfprop}
\end{align}
where $v_0$ is the constant particle velocity or activity.  
$\{\xi_T^i, \xi_r^i\}$ are Gaussian white noises, independent for the different particles $i$, with zero mean $\langle\xi_T^i\rangle =\langle\xi_r^i\rangle = 0$ and correlations $ \langle\xi_T^i(t) \xi_T^i(t')\rangle = 2 D_T\delta(t-t')$, $\langle\xi_r^i(t) \xi_r^i(t')\rangle =2D_R\delta(t-t')$, so that $D_T, D_R$ are the translational and rotational diffusivities, respectively. 
 $\hat{\bf n}(\theta) = (\cos \theta, \sin \theta)$ is the orientation unit 
vector of the particle, giving the 
self-propelling velocity as  $v_0 \hat{\bf n}(\theta)$. The particle motion described by Eqs.\,\eqref{eq:selfprop1}-\eqref{eq:selfprop} fits perfectly into the continuous-limit of the framework developed in the previous section: The translational diffusion and the self-propulsion term would arise from the kind of hopping rates coupled to orientation that have been introduced above, and the rotational diffusion corresponds to the assumed isotropic tumbling. 

Besides the motion, the ABB model in \cite{v3gm-vhry} assumes that particles undergo a birth-death dynamics with a constant death rate $d$ for each particle, and a neighborhood-dependent reproduction rate for each particle $i$ given by $p_i = \textrm{max}(0,b - c N_i^R)$, where $N_i^R$ is the number of particles at distance smaller than $R$ from particle $i$. Thus the 
birth rate of a particle diminishes with the number 
of neighbouring particles in a linear manner. Since a birth rate should be positive the $\textrm{max}$ function imposes the value zero when the linear dependence would lead to negative values. When a reproduction event occurs, the new particle inherits initially the same position and the same angular orientation as the parent\cite{ehgclopezpre}. 

The coupling of the  birth-death-diffusion dynamics (identical to that of the ND model) with active self-propulsion leads to very interesting behavior. In \cite{v3gm-vhry} four different regimes were identified for this particle model, by changing the strength of self-propulsion and the
translational diffusivities: 
i) For small $v_0$ and $D_T$, the $PC$-phase (for \emph{periodic clustered}) was obtained. This is a phase in which clusters of particles form and become arranged with hexagonal order, remaining in place with no global
direction of motion. This is a state similar to the pattern observed
in the ND model without activity.  
ii) For small $D_T$ and large $v_0$ the $SSF$ phase (for \emph{spatially structured flocking}) is obtained. Here particle clusters also appear and 
arrange hexagonally, but now they migrate collectively in a common direction, showing a clear orientational order. This spatially structured flocking
appears without any explicit alignment interaction. iii) For large $D_T$ and $v_0$ a flocking ($F$) phase occurs. It shows
no clear spatial ordering, but exhibits scattered clusters with similar
self-propulsion directions, surrounded by empty regions.
iv) Finally, for large $D_T$ and small $v_0$, a fully disordered ($D$)
phase occurs, with no preferred orientation or travelling direction, and no  spatial ordering.  Our aim here is to recover these phases within our continuum description to get a deeper insight in their causes, besides
checking the flexibility and power of our generalized lattice model defined in Sect. \ref{Sec:Model}.

We proceed to identify the different demographic rates of the ABB in the notation of our lattice-model framework. First, the death rate per particle is a constant $d~G^D_{\bx\theta}=d$. To avoid excessive complication of our continuous modelling approach we will not explicitly impose the positivity of the birth rate, \textit{i.e.} we will use $p_i = b - c N_i^R$ for the reproduction rate of particle $i$. When doing specific simulations, we will take parameter values for which effectively this last rate does not get negative values during evolution. Thus, in the notation of our lattice model, the birth rate would read 

\begin{align}
b~G^B_{\bx\by \theta\beta}(\bn) = \delta_{\bx\by}\delta_{\theta\beta} \left(b - c \LARGE\sum_{\delta,\bz\in \mathcal{R}(\bx)} n_\bz^{\delta} \right)\ ,
\label{eq:discretebirth}
\end{align}
where the sum is over all orientations $\delta$ and over the locations $\bz\in \mathcal{R}(\bx)$ which are within a distance $R$ from $\bx$. The Kronecker deltas implement the fact that the newborn particle inherits the spatial location and angular orientation of the parent. The corresponding continuous-limit rate is 
\begin{align}
b~\frac{G^B_{\bx\by \theta\beta}(\bn)}{\Delta\theta(\Delta x)^d} &\rightarrow 
b~G^B(\bx,\by,\theta,\beta)=\delta(\bx-\by)\delta(\theta-\beta) \times \nonumber \\ & \left(b - c \int d\delta \int_{|\bz-\bx|<R} d\bz ~ \rho(\bz,\delta,t) \right)\ .
\label{eq:continuousbirth}
\end{align}

This is the birth term to be inserted in Eq.(\ref{eq:final_result}). The ABB does not include an explicit pair competition interaction, but as commented above and in Appendix \ref{sec:app_masterEq}, the part in (\ref{eq:discretebirth}) or (\ref{eq:continuousbirth}) which is linearly dependent on $\bn$ can be interpreted (not including the delta functions) as a kind of 
competitive interaction, since it has a negative impact on the population at a location when the population is high in a neighborhood. In this reinterpretation, the birth rate $G^B$ will consist only on a local and  constant part independent of the neighborhood. Thus, decomposing it in the spatial and angular part, $bG^B=bS^B\Phi^B$, we find
\begin{align}
		&\Phi^B(\theta-\theta') = \delta(\theta -\theta')\,, \label{eq:Kb_flocking}
		\\
       &S^B(\bx-\by) = \delta(\bx-\by)  \,.\label{eq:Sb_flocking}
\end{align}

The $\bn$-dependent part in (\ref{eq:continuousbirth}) leads to a competition kernel $\gamma G^C = \gamma S^C\Phi^C$, with 
\begin{align}
        &\Phi^C(\theta-\theta') = 1 \, ,  \label{eq:Kc_flocking}
		\\
		&S^C(\bx-\by) = \begin{cases}
			\dfrac{1}{h}, & |\bx-\by| \le R, \\[6pt]
			           0, & |\bx-\by| > R.
		\end{cases}\label{eq:Sc_flocking}
\end{align}

We normalize the top-hat kernel $S^C$ in Eq.\,\eqref{eq:Sc_flocking} so that its spatial integral becomes unity: $\int d{\bf x}\; S^C(\bx-\by)  = 1$. Thus, in 1D, $h = 2R$ and in 2D, $h = \pi R^2$. When introduced in Eq.\,\eqref{eq:final_result}, the above expressions (\ref{eq:Kb_flocking})-(\ref{eq:Sc_flocking}) recover the effect of introducing Eq.\,\eqref{eq:continuousbirth} if $\gamma=h~c$.

The kernels $\Phi^B$ and $S^B$, Eqs.\,\eqref{eq:Kb_flocking} and \eqref{eq:Sb_flocking}, implement a birth process where the offspring inherits the exact spatial coordinates and orientation of the parent. In contrast, the orientational kernel $\Phi^C$ in Eq.\,\eqref{eq:Kc_flocking} is a constant, implying that competition is isotropic. It occurs with equal intensity between all individuals regardless of their relative orientation. Thus, the model has no explicit orientation-dependent interaction. Finally, the top-hat kernel $S^C$, Eq.\,\eqref{eq:Sc_flocking}, defines a finite interaction range $R$ within which competition is uniformly distributed, effectively imposing a local carrying capacity. This 
nonlocal competitive interaction is the key 
mechanism \cite{ehgclopezpre} responsible, at adequate parameter values,  
for the emergence of spatially periodic ordering (an hexagonal
pattern of particle clusters in 2D) in the absence of self-propulsion. We will see that the same type of patterns appear in this active case, where the spatial ordering will play a relevant role.

%
	

To use the density-polarization formulation given by Eqs.\,\eqref{eq:rho_final} and (\ref{eq:P_final}) we should calculate the contributions $\Delta_n^{B,C}$ from Eqs.\,\eqref{eq:DeltaB} and (\ref{eq:DeltaC}). 
Since the birth angular kernel $\Phi^B$ is local in $\theta$ 
and the competition angular kernel $\Phi^C$ is isotropic, the
birth and competition contributions $\Delta_n^{B,C}$ depend on $\rho_n$ and
$\rho_0$ alone for every $n$, with no further mode coupling, and are
therefore completely unaffected by the choice of closure:
\begin{equation}
	\Delta_n^B = b\,\rho_n , \qquad
	\Delta_n^C = -\gamma\,\rho_n\,(S^C*\rho_0).
\end{equation}

Substituting these into Eqs.\,\eqref{eq:rho_final}--\eqref{eq:P_final}
gives the continuous, von-Mises-closed mean-field description of the
ABB model:
\begin{align}
	\partial_t \rho &= D_T\boldsymbol\nabla^2\rho - v_0\boldsymbol\nabla\cdot\bP
	+ (b-d)\,\rho - \gamma\,\rho\,(S^C*\rho),
	\label{eq:rho_ABB_final}\\[4pt]
	\partial_t \bP &= D_T\boldsymbol\nabla^2\bP - D_R\bP- \frac{v_0}{2}\boldsymbol\nabla\rho
	- {v_0}\boldsymbol\nabla\cdot {\bf Q}
 \nonumber\\&\quad	+ (b-d)\,\bP - \gamma\,\bP\,(S^C*\rho),
	\label{eq:P_ABB_final}
\end{align}
with ${\bf Q}$ given by Eqs.\,\eqref{eq:T_tensor} and \eqref{eq:Bkappa} and $\kappa(\bx,t)=\mathcal A^{-1}(|\bP(\bx,t)|/\rho(\bx,t))$ obtained
from Eq.\,\eqref{eq:order_param}.
The term $-{v_0}\boldsymbol\nabla\cdot {\bf Q}$ in Eq.\,\eqref{eq:P_ABB_final},
vanishes only in the strict $\kappa\to 0$ limit (meaning no polar order) but otherwise remains a nonlinear
correction throughout.

To better understand the dynamics of the coupled 
set of equations (\ref{eq:rho_ABB_final}) and (\ref{eq:P_ABB_final}) it is useful to write them in 
dimensionless form. For that we introduce rescaled variables as
	\begin{align}
& \bx = R \bx' \ , t = \tau t'\ ,   
\rho = \rho_* \rho'
 \ , \bP = \rho_*\bp\ , 
{\bf Q} = \rho_* {\bf Q}' \ ,
	\end{align}
from which {
\begin{align}
{\bf Q}' =\rho' \mathcal B(\kappa)
	\Big(\frac{p_i p_j}{|\bp|^2} - \frac{1}{2}\delta_{ij}\Big) \ . 
\end{align}
}
The scaling time is $\tau = R^2/D_T$, and density, polarization and the tensor ${\bf Q}$ are rescaled by $\rho_* = r/\gamma$, where we have introduced the net growth rate $r = b - d$. In this way we have only three control parameters in the model (besides system size), namely the normalized net growth rate $\mu$, the P\'eclet number $\mathrm{Pe}$ and the normalized rotational diffusion $\overline{D_r}$: 
	\begin{align}
		\mathrm{Pe} = \frac{v_0 R}{D_T}, \quad \overline{D_r} = D_R \tau ,\quad \mu= r \tau.
	\end{align}
To simplify the notation, from now on we will omit the primes in the scaled variables and drop the index from the spatial kernel, $S^C = S$.
	In this notation, the dimensionless equations of motion are 
{
	\begin{align}
		&\partial_t \rho(\bx,t) = \boldsymbol{\nabla}^2 \rho(\bx,t) - \mathrm{Pe}\, \boldsymbol{\nabla}\cdot {\bp}(\bx,t) 
		\nonumber \\
		&\qquad\qquad\qquad +\mu\rho(\bx,t)\bigg( 1 - (S* \rho) (\bx,t)\bigg)\; \label{eq:rho_admensional},   
		\\
		&\partial_t \bp(\bx,t) = \boldsymbol{\nabla}^2 \bp(\bx,t)- \overline{D_r} \bp(\bx,t) - \frac{\mathrm{Pe}}{2} \boldsymbol{\nabla} \rho(\bx,t) \nonumber\\
		&\qquad -\mathrm{Pe} \boldsymbol{\nabla} \cdot {\bf Q}[\rho,\bp]+\mu \bp(\bx,t)  \bigg( 1 - (S* \rho)(\bx,t) \bigg) . \label{eq:pol_admensional} 
	\end{align}

A first trivial solution of this set of equations is the steady and homogeneous state representing the complete absence of particles: $\rho=0$, $\mathbf{p}=\mathbf{0}$. If $\mu>0$ this state is linearly unstable. Then we go further and look for spatially uniform solutions (\textit{i.e.} we set to zero all terms containing spatial derivatives) for $\mu>0$. In this case, Eq.\,\eqref{eq:rho_admensional} becomes a logistic equation for which the solution at long times approaches the constant value $\bar\rho$ given by 
	\[
	\bar\rho = \frac{1}{S_0} = 1 \ .
	\]
We have used $S_0 \equiv \int S(\mathbf x)\,d\mathbf x = 1$. From this, 
the homogeneous solution to  Eq.\,\eqref{eq:pol_admensional} decays to 
\[ 
\mathbf{p}_0=\mathbf{0} \  
	\]
(from which also the tensor ${\bf Q}$ vanishes). This homogeneous and steady solution represents a uniform distribution of particles in space, with absence of mean polar ordering everywhere. Thus, maintaining a non-vanishing polarization at long times requires some degree of inhomogeneity. To understand better the impact of inhomogeneities on the dynamics of configurations close to the steady homogeneous solutions we perform a linear stability analysis in the next section. 

\subsection{Linear stability}
	We denote by $\widetilde S(\mathbf k)$ the Fourier transform of the convolution kernel $S(\mathbf x - \mathbf y)$. Expanding the solutions of  both field equations to first order around the non trivial steady homogeneous solution $\bar\rho=1$ and $\bp_0={\bf 0}$:  
\begin{align}
	\rho(\bx,t)=1 + \delta\rho(\bx,t),\qquad
	\bp(\bx,t)=\delta\bp(\bx,t) \ ,
\end{align}
the nonlocal reaction term linearizes as
\begin{align}
	\mu\,\rho(1-S*\rho)\Big|_{\rho=1+\delta\rho}
	\approx -\mu \,(S*\delta\rho) \,.
\end{align}
Thus the linearized equations become 
	\begin{align}
		\partial_t\,\delta\rho &= \nabla^2\delta\rho - \mathrm{Pe}\,\nabla\!\cdot\!\delta\mathbf p \;-\; \mu \,(S*\delta\rho),
		\label{lin:rho}\\[4pt]
		\partial_t\,\delta\mathbf p &= \nabla^2\delta\mathbf p - \overline{D_r}\,\delta\mathbf p - \tfrac{\mathrm{Pe}}{2}\nabla\delta\rho .
		\label{lin:P}
	\end{align}
Note that the second-order angular Fourier component contained in the tensor ${\bf Q}$ has no effect on this linear analysis. Introducing the Fourier modes {\(\delta \mathcal{F}(\bx,t) \propto \delta \mathcal{F}_\bk e^{\lambda t + i\mathbf k\cdot\mathbf x}\), where $\mathcal{F}$} is either $\rho$ or $\bp$, the linear algebraic system for each Fourier mode becomes
	\begin{align}
		\lambda\,\delta\rho_{\mathbf k} &= -\big(k^2 + \mu\,\widetilde S(k)\big)\,\delta\rho_{\mathbf k} \;-\; i\,\mathrm{Pe}\,\mathbf k\!\cdot\!\delta\mathbf p_{\mathbf k} \ ,
		\label{four:rho}\\[4pt]
		\lambda\,\delta\mathbf p_{\mathbf k} &= -\big(k^2 + \overline{D_r}\big)\,\delta\mathbf p_{\mathbf k} \;-\; i\,\tfrac{\mathrm{Pe}}{2}\,\mathbf k\,\delta\rho_{\mathbf k}\ ,
		\label{four:P}
	\end{align}
where \(k=|\mathbf k|\). It is convenient to decompose \(\delta\mathbf p_{\mathbf k}\) into longitudinal and transverse components to \(\mathbf k\). Let \(p_{\mathbf k}\) be the longitudinal component, \(p_{\mathbf k} = {\mathbf k}\!\cdot\!\delta\mathbf p_{\mathbf k}\). The transverse components decouple and have eigenvalue \(\lambda = -(k^2+\overline{D_r})\). Since this is negative, the homogeneous state is always stable against this type of perturbations. The linear system for \((\delta\rho_{\mathbf k},\,p_{\mathbf k})\) is
{
	\begin{align}
	 ({\bf J}- \lambda \mathbb{I} )
        \begin{pmatrix}
			\delta\rho_{\mathbf k}
			\\
			\delta p_{\mathbf k}
		\end{pmatrix} 
    = 
       \begin{pmatrix}
			0
			\\
			0
		\end{pmatrix},
	\end{align}
}
	with the Jacobian matrix
	\begin{align}
	{\bf J}= \begin{pmatrix}
	&-a(k)\ & -\; i\,\mathrm{Pe}\,k,
	\\
	 &-\; i\,\tfrac{\mathrm{Pe}}{2}\,k&-d(k)
	\end{pmatrix}  .
		\end{align}
We have defined
	\[
	a(k) \equiv k^2 + \mu\,\widetilde S(k),\qquad d(k)\equiv k^2 + \overline{D_r}.
	\]
	
	The characteristic equation for the eigenvalues \(\lambda\) is:
	\begin{equation}\label{eq:char}
		\lambda^\pm = \frac{\mathrm{tr} \;[{\bf J}] }{2} \pm \frac{1}{2}\sqrt{ (\mathrm{tr} \;[{\bf J}] )^2-4\;\mathrm{det}\;[{\bf J}  ]}\;,
	\end{equation}
where 
	\begin{align}
		\mathrm{tr} \;[{\bf J}] = -(a(k)+d(k))\;, 
   \end{align}
   and 
   \begin{align}
\mathrm{det}\;[{\bf J}  ] = 	a(k)\,d(k) + \frac{\mathrm{Pe}^2 k^2}{2} \ .
	\end{align}
The uniform state is linearly unstable at wavenumber \(\bk\) if \(\mathrm{Re}\lbrack\lambda(k)\rbrack>0\) for the largest root of \eqref{eq:char}.

	A necessary condition that guarantees a stable homogeneous state is that  \(\{\mathrm{tr}[{\bf J}] <0 \ \textrm{and}\   \mathrm{det}[{\bf J}  ]>0, \forall~ \bk \}\)  \cite{crossPatternFormationDynamics2009,crossPatternFormationOutside1993}. This stability can be broken in two different ways. The first type of instability, occurring when for some $\bk$, $\mathrm{det}[{\bf J}]$ becomes negative while still $\mathrm{tr}[{\bf J}] <0$, is a stationary instability, also known as type $\mathrm{I}_s$, or Turing instability \cite{crossPatternFormationDynamics2009,crossPatternFormationOutside1993}. In this case the unstable eigenvalue is real, and typically the instability gives rise to a periodic steady configuration of density and polarization. 
This state has the characteristics of the periodic $PC$ phase of the particle ABB model, and this linear stability analysis illustrates its dynamical origin and gives an analytical prediction for its onset. 
  
Substituting \(a(k)\) and \(d(k)\), the Turing instability condition $\mathrm{det}[{\bf J]}<0$ becomes
	\begin{align}
		\mu\, \widetilde S(k) &\;<\; -  \left(
k^2 + \dfrac{\mathrm{Pe}^2}{2}\dfrac{k^2}{k^2 + \overline{D_r}} \right) \ .
		\label{Turing:cond}
	\end{align}
This condition can only be satisfied if for some values of $\bk$ we have $\widetilde S(k)<0$ (as is indeed the case for our choice of kernel (\ref{eq:Sc_flocking})).
This condition of negativity of $\widetilde S(k)$ 
for some wavenumbers is also needed for pattern 
formation in the absence 
of active motion \cite{lopezehgphysD}. In fact 
Eq. (\ref{Turing:cond}) reduces to the instability condition for the passive ND 
Brownian bug model when $\textrm{Pe}=0$. The critical 
value of $\mu$ leading to instability increases with $\textrm{Pe}$. This critical $\mu_c(\textrm{Pe})$ and the associated $k_c$ can be obtained by finding the smallest $\mu$ for which $\mathrm{det}[{\bf J]}=0$. 

The other type of instability, an oscillatory $\mathrm{I}_o$ transition (or Hopf with finite wavenumber), occurs when $\mathrm{tr} \;[{\bf J}]$ becomes positive, whereas still $\mathrm{det}\;[{\bf J}  ]>0$. Typically, this leads to the appearance of travelling waves in the system, a state with the characteristics of $SFF$-phase of the particle ABB model. Again, our linear stability analysis allows an analytical prediction for the onset of this phase: it will appear for $\mu>\mu_c$, where 
\begin{align}
	\mu_c = - \frac{2 k_c^2 +\overline{D}_r}{\widetilde S(k_c)}, \label{eq:hopf_condition}
\end{align}
and $k_c$ can be obtained from the solution of $\mathrm{tr} \;[{\bf J}]=0$ with the minimum $\mu$. This instability needs again that $\widetilde S(k_c)<0$  for the wavenumber $k_c$ leading to maximum real part of the eigenvalue. It leads to travelling waves in the system, with phase velocity $\pm \mathrm{Im}(\lambda(k_c))/k_c$. Note that Eq.\,(\ref{eq:hopf_condition}) is independent of $\textrm{Pe}$, although it only identifies the dominant instability as far as $(\mathrm{tr} \;[{\bf J}] )^2-4\;\mathrm{det}\;[{\bf J}  ]<0$. The condition $\mathrm{tr} \;[{\bf J}] =  \mathrm{det}\;[{\bf J}  ]=0$ locates a Turing-Hopf point.

The linear-stability relationship Eq.\,\eqref{Turing:cond} 
identifies modes and thresholds for 
the \emph{onset} of pattern formation, but it 
does not predict nonlinear 
outcomes. 
The same can be said for the Hopf condition, Eq.\,\eqref{eq:hopf_condition}, that determines 
the region where oscillations start to grow and propagate in the system.
In the next section we perform numerical simulations 
to get deeper insight into the phenomenology of our continuum description beyond instability onset.

	\subsection{Numerical Simulations}

	We present numerical simulations of our 
	 continuum description of the ABB model, 
	 informed by the linear stability analysis 
	 of the preceding section.
	  We first consider a 
	 quasi-one-dimensional case, which is computationally less demanding, 
	 and then we move to two spatial dimensions, where 
	 direct comparison with the original 
	 ABB particle model becomes possible.
	For both cases we implement a standard pseudospectral algorithm integration for the nonlocal interactions \cite{mendesSpectralMethods2019,silvanoFlowSpatialStructure2025}. The code and all the details can be found in the  \href{https://github.com/Jaegg3rNat/ABP_Flocking}{GitHub} repository \cite{Jaegg3rNat_ABP_Flocking}.

	\subsubsection{Simulations in a quasi-one-dimensional case}
	
	In order to observe the instability 
	transitions and the spatial structures 
	of the previous section we resort to numerical simulations. 
	Note that we have derived our general 
	model in the two-dimensional situation, but numerical simulations and analytical understanding become simpler in 1D.  As we mentioned, the derivations of the general model 
	above can be repeated in 1D, just considering that 
	there are only two possible orientations, say $M=\pm 1$\cite{demaerelActiveProcessesOne2018}. 
	But the lack of continuity between the
	 orientional $M$-states would lead to a phenomenology 
	 quite different from the two-dimensional case, which is our focus for 
	 a closer comparison with \cite{v3gm-vhry}.
	 Thus, we define a type of quasi-one-dimensional 
	 model which still contains the essential 2D phenomena 
	 but is easier to analyze computationally and theoretically.
	  Essentially, we replace the active self-propulsion vector $v_0\hat\bn(\theta)$ by its projection
	  on the $x$ direction, so that the active velocity is $v_0(cos \theta, 0)$. This would
	  correspond to an active system constrained to move only in one direction (for 
	  example through a narrow capillary). The orientation vector can still be arbitrarily oriented,
	  but only its $x$-component manifests into self-propulsion. If in addition we assume that 
	  initially the angular distribution is symmetric around the horizontal axis, so that $p_y=0$, 
	  this symmetry turns out to be preserved at all times. We have $\bp=(p_x,0)$ and  
	  the equations of motion for this quasi-one-dimensional case simplify to (see details in Appendix C): 

	The equations of motion for this quasi-one-dimensional case simplify to
	\begin{align}
		&\partial_t \rho = \partial_{ x}^2 \rho - \mathrm{Pe}\; \partial_{ x} p_{ x} +\mu\rho\bigg( 1 - (S* \rho) \bigg)  , \label{eq:1drho}
		\\
		&\partial_t p_{ x} = \partial_{ x}^2 p_{ x}- \overline{D_r} p_{ x} - \frac{\mathrm{Pe}}{2}\partial_{ x}\bigg\lbrack  \rho + {\rho\mathcal B(\kappa)}  \bigg\rbrack \nonumber \\ &\qquad \qquad+\mu  p_{ x}  \bigg( 1 - (S* \rho) \bigg) \label{eq:1dPx}   .
	\end{align}
	We use the normalized nonlocal top-hat kernel of Eq.\,\eqref{eq:Sc_flocking} in 1D (remember we are using units such that $R=1$),
	\begin{align}
	S({ x}) = \begin{cases}
		\dfrac{1}{2}, & |{ x}| \le 1, \\[6pt]
		0, & |{ x}| > 1,
	\end{cases}
	\end{align}
whose Fourier transform is,
	\begin{align}
		\tilde{S}(k) = \frac{\sin (k)}{k}.
	\end{align}
	Note that $\tilde S(k)$ takes negative values for some $k$.
The polarization $p_x$ is a continuous field, 
where its sign reveals the local mean 
direction of motion, to the left ($-$) or to the right ($+$).
Performing the numerical simulations of 
Eqs.\,\eqref{eq:1drho} and \eqref{eq:1dPx}, starting from the homogeneous solution slightly perturbed by noise, we identify 
different regimes of collective motion depending 
on the region of parameters in the phase space.
The different dynamical regimes of motion approached at long times, are represented in Fig \ref{fig:phases1d}:
(a) and (e) homogeneous density and zero global polarization. This is analogous to the disordered $D$-phase in the ABB model. (b) and (f) show non-homogeneous density coupled to non-zero local values of polarization. Both forming spatial periodic patterns but with zero global polarization (see definition of the order parameter below, Eq. (\ref{eq:order_parameter})). This is a state analogous to the periodic clustered ($PC$ phase) state of the particle ABB model. 
(c)-(g) show non-homogeneous (periodic) density and local polarization, and
non-zero global polarization, leading to a global drifting of the whole pattern. This is analogous to the $SFF$ phase in the ABB.
	
Fig.\,\ref{fig:phase_diagram1d} displays 
	  the phase space spanned by the parameters $\mu$ and $\mathrm{Pe}$, together with the lines of linear instability, 
	  for fixed dimensionless tumbling rate $\overline{D_r} = 0.7$. 
	Fig.\,\ref{fig:phase_diagram1d}a) displays the spatially averaged  density of the system at long times. 
\[
	\langle\rho\rangle = \frac{1}{L}\int dx \rho(x, \infty),
\] 
which has unit value when the configurations are homogeneous. We see an increase of $\langle\rho\rangle$ from the homogeneous unit value when the linear instability lines are crossed, which serves as an indicator of the onset of pattern formation in the system. 

Fig.\,\ref{fig:phase_diagram1d}b) displays the global polarization, \textit{i.e.},
 the order parameter ${\psi}$  of the system, defined as
	\begin{align}
		 \psi 
		 = \frac{1}{L}\bigg| \int  \frac{p_x(x,\infty )}{\rho(x,\infty )}  dx \bigg|\;\label{eq:order_parameter}.
	\end{align} 
The analytical prediction for the Turing instability line, Eq.\,(\ref{Turing:cond}), is depicted in the dashed gray line in Fig.\,\ref{fig:phase_diagram1d} that separates the phase space 
into the region where the homogeneous solution is stable ($\langle\rho\rangle = 1$, below the curve) 
and the spatially periodic patterns phase ($\langle\rho\rangle>1$, above the curve). 
In the homogeneous state there is no flocking,
 corresponding to a disordered state of the particle velocities
  with $\psi=0$ (the $D$-phase) 
  behaviour (Figs.\,\ref{fig:phases1d}a and  \ref{fig:phases1d}e, \href{https://cloud.ifisc.uib-csic.es/nextcloud/index.php/s/tFEFDPxWR3jrExb}{SM Video 1}~\cite{SupplementalMaterial}). 
  For low $\mu$ and/or low $\mathrm{Pe}$ the numerical 
  simulation correctly captures the development 
  of the Turing instability: when starting from
   the homogeneous nonzero solution slightly perturbed 
   with noise, a steady spatially periodic state for the density and for the polarization is reached (Figs.\,\ref{fig:phases1d}b, and  \ref{fig:phases1d}f, \href{https://cloud.ifisc.uib-csic.es/nextcloud/index.php/s/iLoWitpXgkjxgWt}{SM Video 2}~\cite{SupplementalMaterial}). The polarization oscillates between negative and positive values, reflecting the tendency of the particles to go left or right before dying when leaving the clusters, but there is no global motion of the whole pattern, as indicated by $\psi\sim 0$. This is analogous to the $PC$-phase in the particle model.
   The onset of pattern formation shifts to higher 
   values of $\mu$ with the increase of $\mathrm{Pe}$ as predicted. 
 
For larger values of $\mu$ and $\mathrm{Pe}$ the patterns form and remain static during a transient, with positive and negative values of $p_x$ symmetrically distributed on the flanks of the clusters. But after this transient they start to drift as
  a whole, with constant velocity, 
  spontaneously breaking the left-right symmetry (meaning that $p_x$ becomes positive or negative on the whole system). This parameter region
  (corresponding to the particle $SFF$-phase)
   of drifting patterns is identified in 
   Fig.\,\ref{fig:phase_diagram1d}b by a large value of the global polarization ($\psi  \sim  1$, see Figs.\,\ref{fig:phases1d}c, \ref{fig:phases1d}g and \href{https://cloud.ifisc.uib-csic.es/nextcloud/index.php/s/QrdyZrcria9SmMo}{SM Video 3}~\cite{SupplementalMaterial}). 
		\begin{figure*}[hbt!]
		\includegraphics[width=\textwidth]{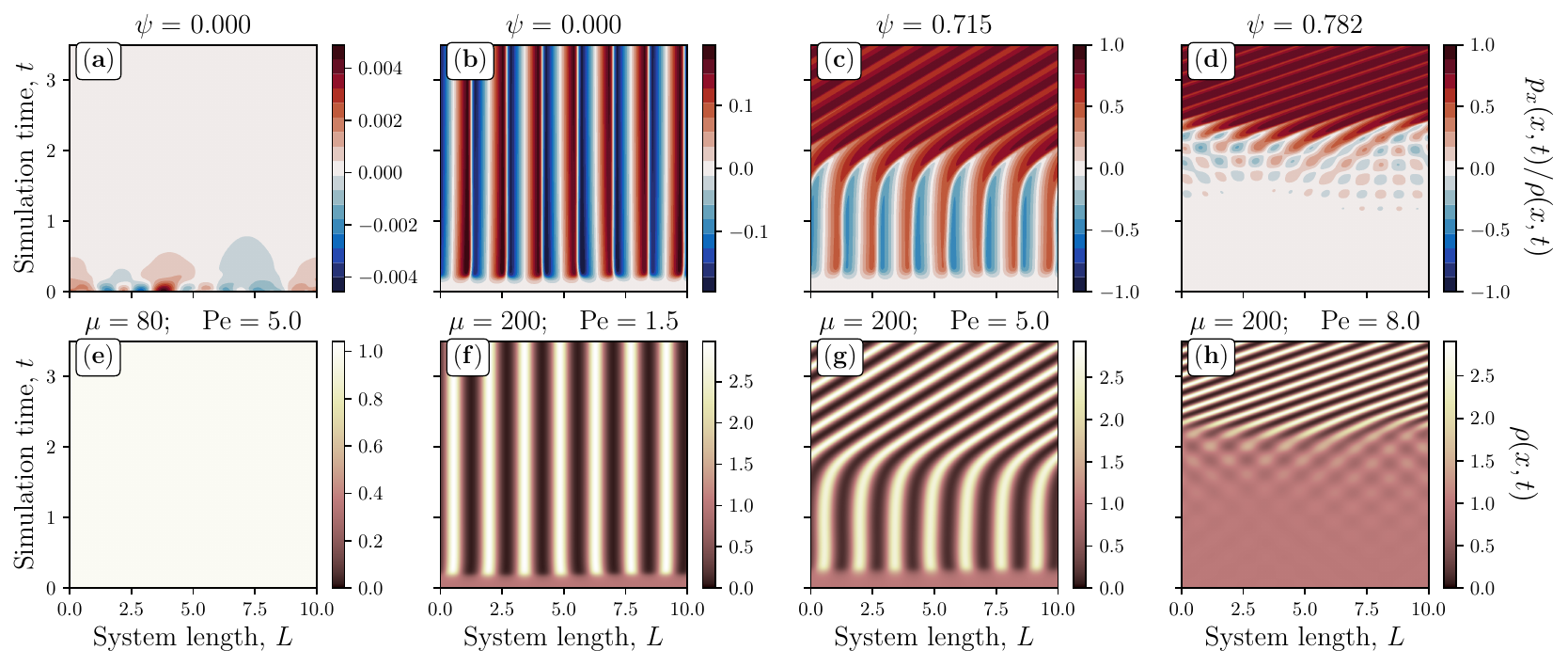}
		\caption{System evolution in the quasi-one-dimensional case, Eqs.\,(\ref{eq:1drho}) and (\ref{eq:1dPx}), for different dynamical regimes.
			Bottom row is the normalized density distribution $\rho(x,t)$. Top row
			is the polarization strength $p_x(x,t)/\rho(x,t)$. We identify, at the
			final simulation time, the order parameter $\psi$ computed from Eq.\,(\ref{eq:order_parameter}). Values of $\mu$ and $\mathrm{Pe}$ are
			indicated in each case. Other parameters: $L=100$ and $\overline{D_r}
			=0.7$. Initial condition is always the homogeneous solution $\rho=1$,
			$p_x=0$ slightly perturbed by noise.  \textbf{a, e)} Evolution relaxing
			towards the stable homogeneous density with null polarization.
			\textbf{b, f)} Formation of stationary patterns with negligible
			orientational order ($\psi \sim 0$). \textbf{c, g)} Pattern formation by
			the Turing instability, but followed by a transition to traveling waves
			(flocking) with large global polarization $\psi$. \textbf{d, h)} Hopf
			instability leading to traveling patterns with large global polarization
			$\psi$.
}
		\label{fig:phases1d}
	\end{figure*}
	
		\begin{figure}[hbt!]
			\includegraphics[width=\columnwidth]{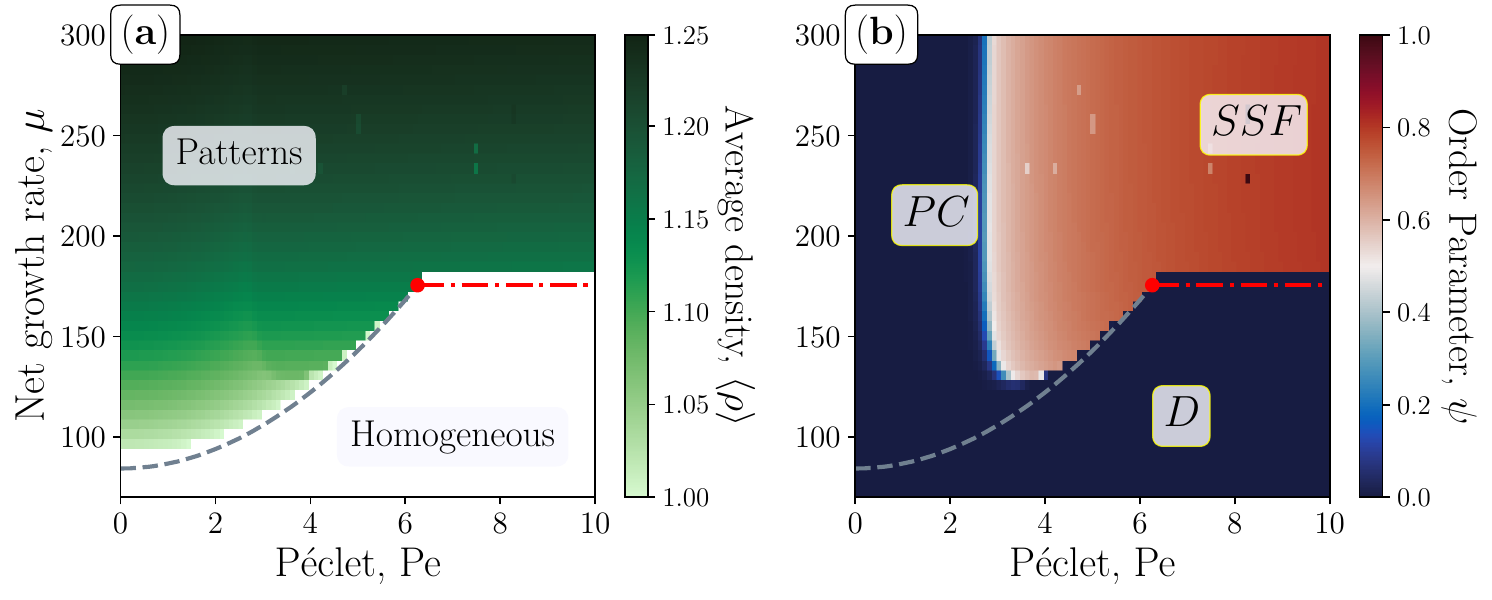}
\caption{Phase Diagram simulated from Eq.\,\eqref{eq:1drho} and \eqref{eq:1dPx}
	in one spatial dimension, at a final time $t=25$ starting from the homogeneous solution slightly
	perturbed by noise. The dashed gray line is the analytical Turing line of
	pattern formation from Eq.\,\eqref{Turing:cond}. The red dash dotted line is the
	analytical Hopf instability line from Eq.\,\eqref{eq:hopf_condition}.
	Color in \textbf{a)} indicates the spatially averaged density
	$\langle\rho\rangle$, where $\langle\rho\rangle = 1$ identifies the
	homogeneous steady state. Color in \textbf{b)} indicates the order
	parameter $\psi$, computed from Eq.\,\eqref{eq:order_parameter}. Other
	parameters: $L =10$ and $\overline{D_r} =0.7$. The different phases
	found are indicated with the labels D, PC and SSF in panel b). The D
	phase gives a homogeneous density, as indicated in panel a). SSF and PC
	display spatial patterns.}
			\label{fig:phase_diagram1d}
	\end{figure}
	
This behaviour is not readily explained by the 
linear instability analysis, and it is produced 
by strong non-linear effects 
   mediating the interplay between the Turing 
   patterns and the travelling waves that appear in the Hopf bifurcation occurring at higher values of $\mu$ and $\textrm{Pe}$. 

The red dash-dotted line in Fig.\ref{fig:phase_diagram1d}a) is the analytical prediction for the Hopf transition, Eq.\eqref{eq:hopf_condition}. Above this line, density/polarization fluctuations propagate in the system in form of periodic waves (Figs.\,\ref{fig:phases1d}d, \ref{fig:phases1d}h, and \href{https://cloud.ifisc.uib-csic.es/nextcloud/index.php/s/xn9Wfb8k6Nj2EfQ}{SM Video 4}~\cite{SupplementalMaterial}).

Thus, we show that
in this continuous-model representation, as in the original particle ABB model, collective motion emerges without any explicit alignment interaction. Rather, the mechanism appears as a coupling between density 
 gradients and polarization: it is
  necessary to have some spatial inhomogeneity.
  From the continuum equations we can gain
  some physical insight on the mechanisms (in particular
  for the most interesting $SFF$-phase).
When active self-propulsion is zero, the equation reduces to a nonlocal 
FKKP equation, in which the clusters are stationary and periodically arranged
\cite{ehgclopezpre}. When activity 
 is included, localized density clusters develop 
 asymmetric profiles of polarization. Through 
 the term $-\frac{\mathrm{Pe}}{2}\;\partial_{ x} \rho$ 
 in the polarization dynamics, these asymmetric 
 density gradients generate a finite polarization at the cluster edges, and there is a tendency to advection via the $-\mathrm{Pe} \;\partial_{x} p_{ x}$ coupling in the density equation.
 Returning (note that this explanation is independent on
 the spatial dimension) to the original particle interpretation of the
  model, this means that particles close to the edges tend to move towards the outside.
   In the regime where $\psi=0$, however, particles leaving the cluster enter a low-density environment in which reproduction cannot compensate for larger death by competition. They therefore disappear before generating sustained transport, so that the center of mass of each cluster remains stationary despite the local outward particle flux. As either $\mu$ or $\mathrm{Pe}$ is increased, the polarization-induced advective coupling becomes sufficiently strong to overcome this balance, producing a persistent displacement of the entire cluster. This corresponds to a drift instability in which the cluster acquires a finite propagation velocity.

Importantly, the nonlocal interaction couples the dynamics of neighboring clusters through the birth and competition terms, making their drift velocities interdependent. As a result, once the drifting state is established, the clusters spontaneously select the same propagation direction and move with a common velocity rather than drifting independently. The resulting state consists of coherently propagating density patterns that resemble flocking at the macroscopic level. Although, at the continuum level, this behavior appears as pattern-induced self-advection, it is the coarse-grained manifestation of the spontaneous polar ordering that emerges in the underlying particle system.

Previous simulations always started from the homogeneous solution perturbed by noise. The transition between homogeneous, stationary patterned, and traveling patterned states exhibits regions of bistability and hysteresis, indicating subcritical behavior of both the Turing and Hopf instabilities. {The corresponding analysis sweeping parameters back and forward is presented in Appendix~\ref{sec:app_hysteresis}.}

	\subsubsection{Two-dimensional case}

The qualitative behavior observed for quasi-one-dimensional dynamics persists in two-dimensional systems. However, the extension to two 
 spatial dimensions introduces important qualitative 
 and quantitative differences due to the vectorial nature of the polarization field $\mathbf{p} = (p_x, p_y)$ capturing the continuous symmetry of particle orientations $\theta \in [0,2\pi)$, which allows for continuous reorientations in space. This continuous symmetry plays a
 role in enhancing collective effects, 
 particularly in the emergence of macroscopic flocking states.
 The computation of the order parameter will be generalized for two dimensions as,
 \begin{align}
 	\psi =\frac{1}{L^2} \bigg|\int {\bf dx} \frac{\bP(\bx, \infty)}{\rho(\bx, \infty)} \bigg|
 	\label{eq:op2}
 \end{align}
The governing equations are given by Eqs.\,\eqref{eq:rho_admensional},\eqref{eq:pol_admensional}.
For the two-dimensional case the top-hat kernel (in units in which $R=1$) is,
\begin{align}
	S({\bf x}) = \begin{cases}
		\dfrac{1}{\pi}, & |{\bf x}| \le 1, \\[6pt]
		0, & |{\bf x}| > 1,
	\end{cases}
\end{align}
and the Fourier transform,\begin{align}
	\tilde S(k) = 2\frac{J_1(k)}{k}.
\end{align}

As in the quasi-one-dimensional case, the system exhibits
 three main dynamical regimes depending on the parameters. We present in Fig.\,\ref{fig:phase_diagram2d} the phase diagram spanned by $\mu$ and $\mathrm{Pe})$, for fixed $\overline{D_r} = 0.7$. We find the following states:
  (i) a homogeneous disordered phase (the $D$-phase in the ABB), 
  (ii) a Turing like pattern forming
   phase without global motion ($PC$-phase), and 
   (iii) a phase of traveling patterns associated 
   with macroscopic flocking ($SFF$-phase).
  In two dimensions, local 
  polarization vectors can continuously 
  reorient and align with neighboring regions 
  through the coupling to density gradients. This facilitates 
  the development of long-range correlations 
  in the polarization field, stabilizing coherent 
  motion at the macroscopic scale. In contrast, 
  in quasi-one-dimensional systems, polarization is restricted to
   a scalar value, which limits the 
   ability of the system to sustain global alignment.

Within the Turing unstable region, for small $\mathrm{Pe}$, the system develops 
spatially periodic density pattern (Fig \ref{fig:phases2d}a, \href{https://cloud.ifisc.uib-csic.es/nextcloud/index.php/s/qSGBEGcJrK7A2wd}{SM Video 5}~\cite{SupplementalMaterial}), 
which in two dimensions takes the form 
of hexagonal structures. These structures are 
accompanied by spatially varying polarization 
fields that globally cancel out, yielding $\psi \approx 0$. As
 in one dimension, these states correspond to 
 stationary patterns resulting from the 
 balance between diffusion, growth, and nonlocal interactions, modified by selfpropulsion indicated by $\mathbf{p}$.
Due to the vectorial nature of $\mathbf{p}$, the polarization 
field in these patterns exhibits topological defect configurations of sources (or asters) in the \textit{PC}-phase. These features have no analogue in the quasi-1D case and play an important role in the dynamical evolution of the system. A detailed study of these structures is left for future work.

At larger values of $\mu$ and $\mathrm{Pe}$, the periodic patterns still form and remain stationary during a transient, but as in the quasi-one-dimensional case they eventually lose stability and transition into traveling states (Fig~\ref{fig:phases2d}b, \href{https://cloud.ifisc.uib-csic.es/nextcloud/index.php/s/BjFGiCGSXLFC6BC}{SM Video 6}~\cite{SupplementalMaterial}), in which the topological defects in the polarization field disappear. This point to nonlinear interactions between Turing modes (spatial structure) and Hopf modes (temporal oscillations). 
In this flocking regime (which in the particle model corresponds to the $SFF$-phase), 
density structures propagate  across the system 
while maintaining their shape, resulting 
in a non-zero global polarization $\psi > 0$. 
Again, the mechanism underlying this 
transition is analogous to the quasi-one-dimensional case. 
The local inhomogeneities, in the form of periodic 
clusters, of the density field generate polarization through the 
$-\frac{\mathrm{Pe}}{2}\boldsymbol\nabla \rho$ 
term, which in turn advects the density via
 the $-\mathrm{Pe}\,\boldsymbol\nabla \cdot \mathbf{p}$ coupling. 
 In two dimensions, this feedback loop can occur
  along multiple directions simultaneously, allowing 
  clusters to not only drift but also reorient collectively.

For $\mu$ and Pe above the Hopf instability (red line in Fig.\ref{fig:phase_diagram2d}) the SSF phase develops directly in 
an oscillatory instability ( \href{https://cloud.ifisc.uib-csic.es/nextcloud/index.php/s/sytto4j7ff58CEa}{SM Video 7}~\cite{SupplementalMaterial}), but also as a secondary instability of the periodic pattern occurring after the Turing instability (\href{https://cloud.ifisc.uib-csic.es/nextcloud/index.php/s/BjFGiCGSXLFC6BC}{SM Video 6}~\cite{SupplementalMaterial}). This points to nonlinear interactions between Turing modes 
 (spatial structure) and Hopf modes (temporal oscillations). In addition to the phenomenology described above, the bistability regimes and hysteresis reported in  Appendix~\ref{sec:app_hysteresis}  for the quasi-one-dimensional case are also present in this two-dimensional situation.

		\begin{figure}[hbt!]
		\includegraphics[width=\columnwidth]{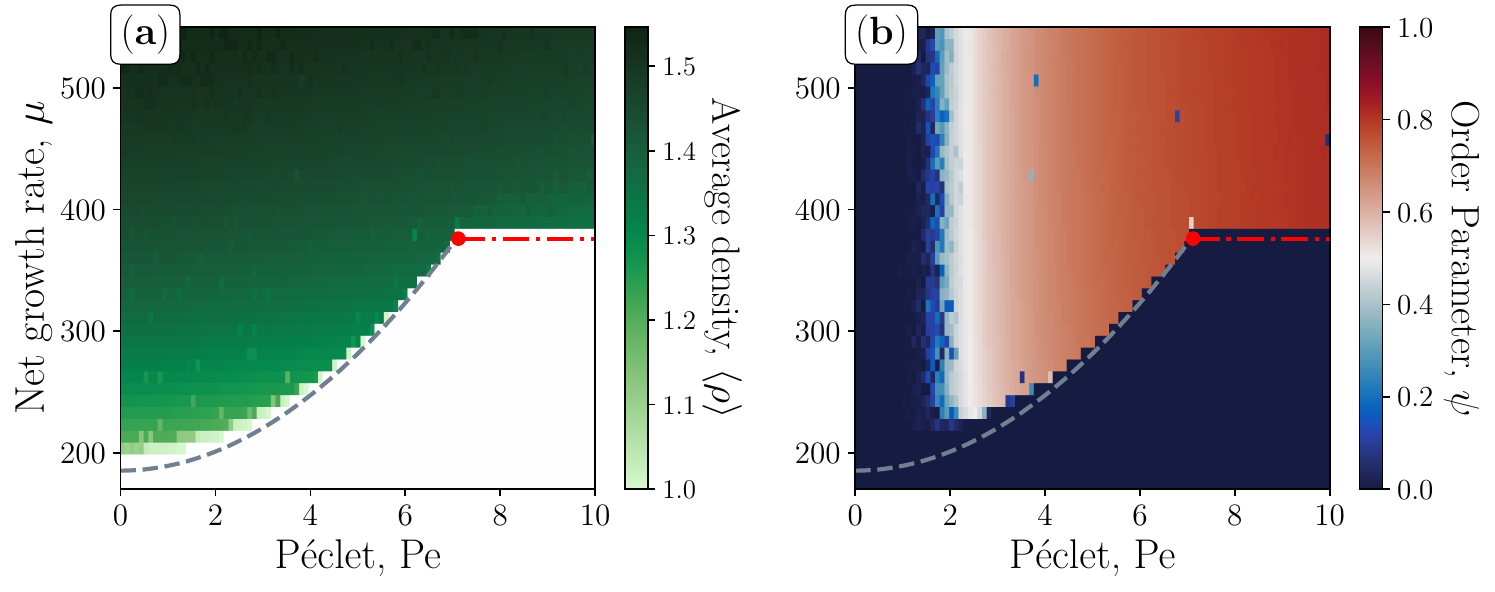}
		\caption{Phase Diagram simulated from
			Eqs.\,\eqref{eq:rho_admensional},\eqref{eq:pol_admensional} in two
			spatial dimension, at a final time $t= 50$ starting from the homogeneous solution slightly
			perturbed by noise. The dashed gray line is the analytical Turing line of
			pattern formation from Eq.\,\eqref{Turing:cond}. The red dash dotted line is the
			analytical Hopf instability line from Eq.\,\eqref{eq:hopf_condition}.
			Color in \textbf{a)} indicates the spatially averaged density
			$\langle\rho\rangle$, where $\langle\rho\rangle = 1$ identifies the
			homogeneous steady state. Color in \textbf{b)} indicates the order
			parameter $\psi$, computed from Eq.\,\eqref{eq:op2}. Other parameters: $L =10$ and $\overline{D_r}
			=0.7$.
			}
		\label{fig:phase_diagram2d}
	\end{figure}
	
		\begin{figure*}[hbt!]
		\includegraphics[width=0.9\textwidth]{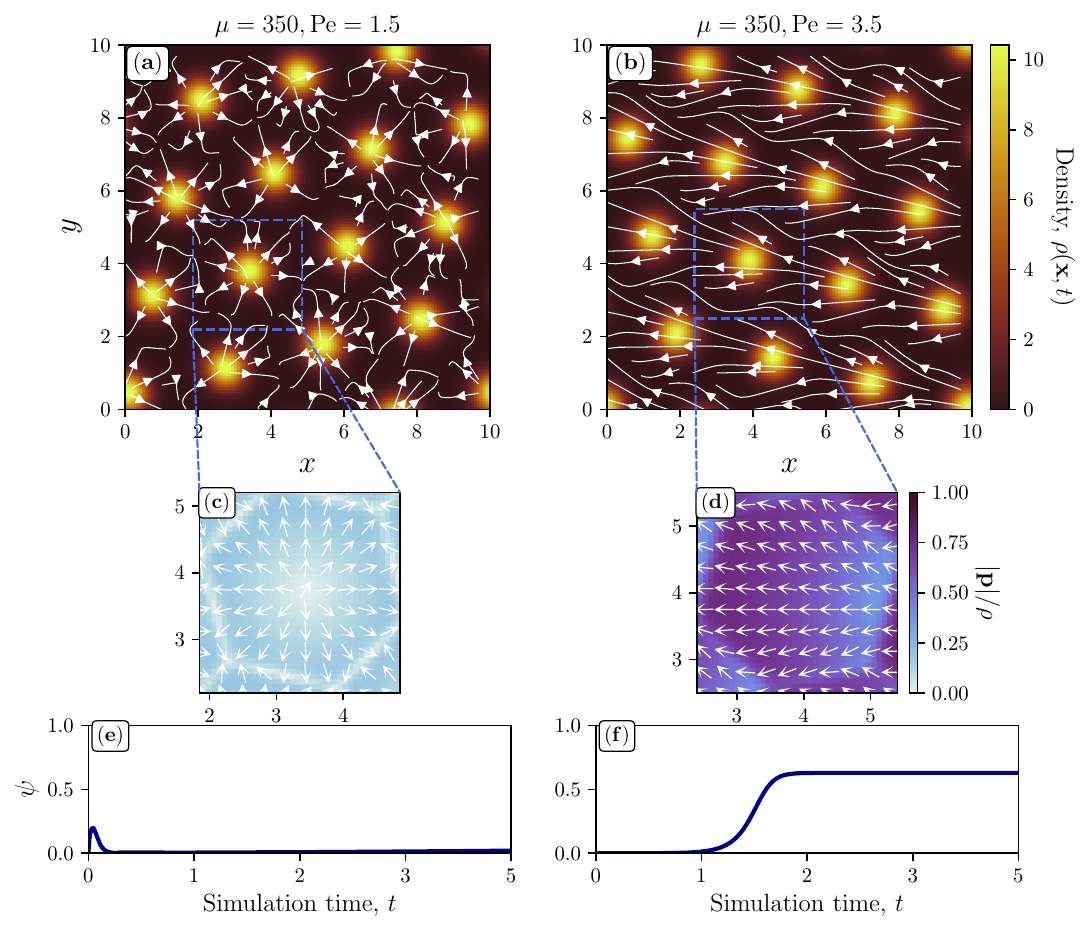}
		\caption{Spatiotemporal patterns from numerical integration of Eqs.\,\eqref{eq:rho_admensional} and \eqref{eq:pol_admensional} in two
			dimensions starting from the homogeneous solution slightly perturbed
			with noise. In all panels $L=10$ and $\overline{D_r} =0.7$. \textbf{(a,
				b)} Density configurations $\rho$ at long times, with polarization
			streamlines in white. \textbf{(a)} Stationary hexagonal clusters
			obtained at $\mu = 350$ and  $\mathrm{Pe} = 1.5$ (see
			\href{https://cloud.ifisc.uib-csic.es/nextcloud/index.php/s/qSGBEGcJrK7A2wd}{SM
				Video 5}~\cite{SupplementalMaterial}) exhibiting in \textbf{(c)} periodic angular orientations with
			small polarization strengths $|\bp|/\rho $. \textbf{(b)} Traveling
			hexagonal clusters obtained at $\mu = 350$ and $\mathrm{Pe} = 3.5$,
			characterized by aligned orientations and high polarization strength
			$|{\bp}|/\rho$ (shown in \textbf{(d)}. This configuration is reached
			after a transient in which stationary clusters such as those in
			\textbf{(a)} form initially, and then break the globally isotropic
			symmetry to develop the traveling (flocking) state (see
			\href{https://cloud.ifisc.uib-csic.es/nextcloud/index.php/s/BjFGiCGSXLFC6BC}{SM
				Video 6}~\cite{SupplementalMaterial}). \textbf{(e, f)} Dynamics of the order parameter $\psi$ ending
			in the configurations shown in \textbf{(a)} and \textbf{(b)}, respectively.}
		\label{fig:phases2d}
	\end{figure*}

	\section{Summary and Conclusions}
       \label{Sec:summary}
       
In this work, we have introduced a generalized
 lattice-based model for active particles 
 subject to reproduction, death, and spatially 
 nonlocal competitive interactions. Starting from 
 a microscopic master equation, we derived a 
 mean-field continuum description that after a suitable closure, consists of 
 coupled evolution equations for the particle density $\rho(\mathbf{x},t)$
and the polarization field $\mathbf{P}(\mathbf{x},t)$. 
This framework provides a systematic 
 approach to the study of proliferating 
 active matter, applicable to a broad class of
  individual-based models with different interaction kernels.
As a concrete application, we applied the 
continuum framework to the Active Brownian Bug  model \cite{v3gm-vhry}, a recently 
introduced model in which flocking 
emerges solely from the interplay between 
self-propulsion and density-dependent 
proliferation dynamics, without any 
explicit alignment interaction. By using the spatial and orientational kernels appropriate to the dynamics in \cite{v3gm-vhry}, 
 the general equations reduce to a 
 tractable pair of coupled equations
  for the density and polarization fields, 
 which were then studied both analytically and numerically.
 
The linear stability analysis of the homogeneous
 steady state revealed two 
 distinct instability mechanisms. The first is a Turing 
  instability leading to steady patterns
  and the  second is an Hopf 
  instability at finite wavenumber, which drives 
  the emergence of traveling density and polarization waves. 
Three main dynamical regimes are observed (both in 1D and 2D): 
a homogeneous disordered phase, a stationary patterned 
phase with no global motion, and a traveling 
pattern phase associated with macroscopic collective motion.
These correspond, respectively, to the $D$, $PC$ and $SFF$ phases
observed in the original particle ABB model in \cite{v3gm-vhry}.
Comparing our results with the particle model in \cite{v3gm-vhry}, 
we notice that an analogous to the $F$-phase (a state with disordered clusters but with a global polar ordering leading to flocking) is absent.
We believe that this phase arises from 
particle density fluctuations, which are needed to sustain polarization, and thus noise terms would be needed in the continuum equations
to properly account for them. In fact we suspect that particle interactions are not really needed to produce such phase, but just the reproductive correlations (newborn having the same location and orientation than parent) that appear in the stochastic description of the particle system \cite{Young2001}. Introducing the corresponding fluctuations in the continuous description is beyond the scope of the present paper, and would be subject of future work. 

In summary, the continuum description introduced here 
 successfully captures most of the main 
 dynamical regimes of the ABB model, and
  provides an interpretation of the onset 
of collective motion: flocking does not arise 
from explicit orientation interactions, but rather from 
a density-polarization coupling that 
generates a drift when spatial structures develop. 
Nonlocal interactions couple the dynamics of neighboring clusters through the birth and competition terms, making their drift velocities interdependent. The analytical linear stability analysis accurately predicts the onset of both the Turing and Hopf instabilities, while numerical continuation through forward and backward parameter sweeps reveals hysteresis and coexistence regions, demonstrating that both transitions exhibit subcritical behavior due to nonlinear effects.

Several directions remain open for future work.
A weakly nonlinear  analysis near the Turing 
 and Hopf bifurcation points would provide a 
more systematic understanding of the transition 
between stationary and traveling patterns.
Also, including fluctuation terms in the continuum equations
would be closer to the particle description and hopefully
recover all the phases observed in the
ABB. 
Extensions of the model to include explicit alignment interactions,
or more complex proliferation rules would 
broaden its applicability to other systems 
of biological and physical interest. 

	\acknowledgments
	
	N.S. and E.H-G. acknowledge 
	funding by the Spanish Ministerio de Ciencia, 
	Innovaci\'on y Universidades
	(MICIU/AEI/10.13039/501100011033) through
	the Maria de Maeztu project CEX2021-001164-M.
	C.L. and E.H.-G. acknowledge Grant No. LAMARCA PID2021-123352OB-C32 funded by MICIU/AEI/10.13039/501100011033 and FEDER, UE.

	\appendix
	\section{Master Equation}\label{sec:app_masterEq}

In this Appendix we will provide the detailed description of the master equation of our microscopic model. The Master Equation derivation from reaction rates is a calculation that can be found in standard stochastic processes books \cite{toralStochasticNumericalMethods2014,gardiner4}. Here, however, we have as ingredients the orientation degree of freedom and the self-propulsion, for which we believe that giving the full description could be beneficial for reproducibility by the interested reader.

As we mentioned in the text, our particles reside in a two dimensional lattice and the orientation angle is a discrete variable in 2D.
The entries $n_\bx^\theta$ in the state vector $\bn$ that describes the system are the number of particles at position ${\bf x}$ and orientation $\theta$.

	\subsection{Master Equation for Birth Process}
	\begin{figure}[h!]
		\includegraphics[width=0.95\columnwidth]{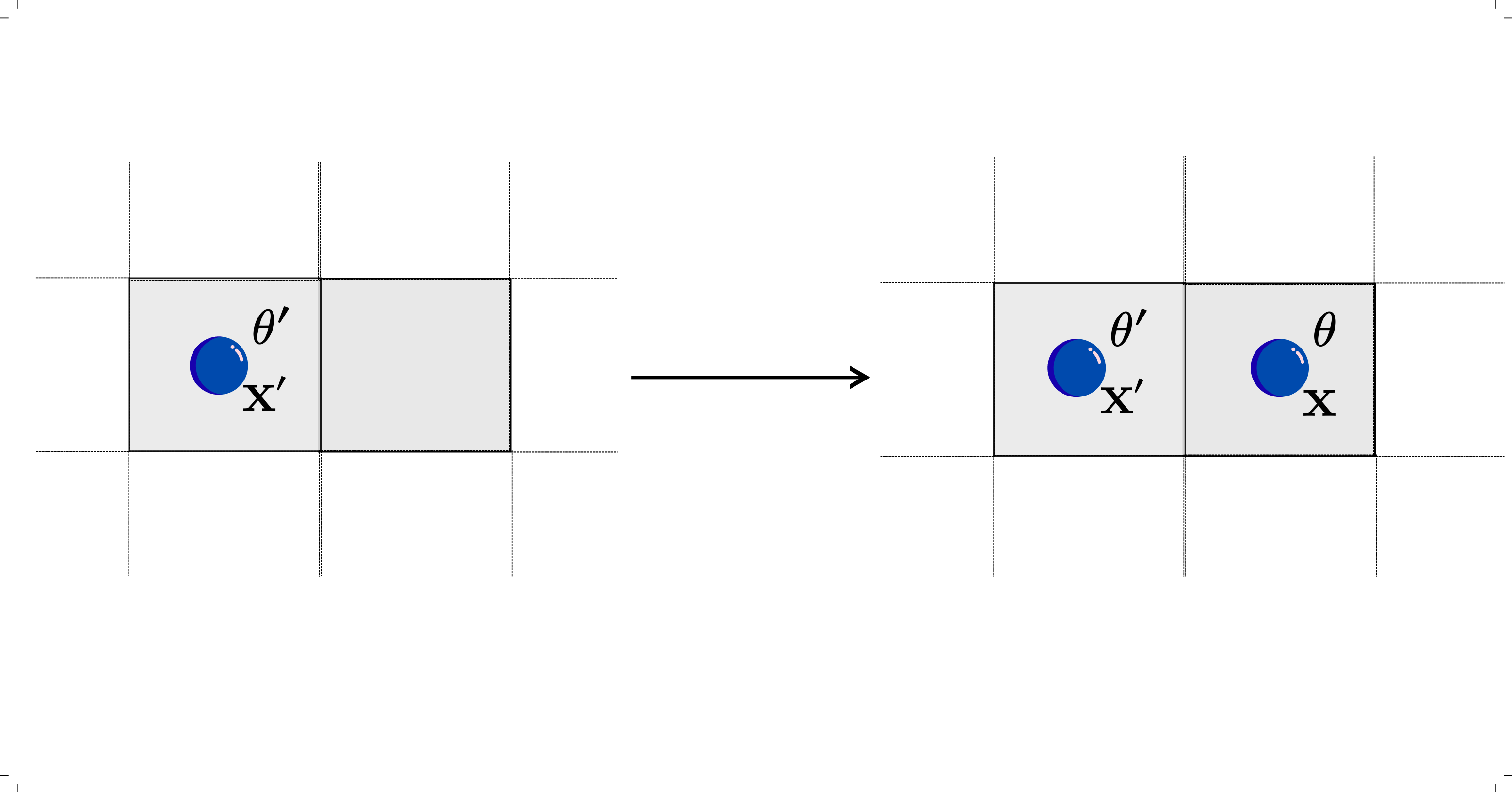}
\caption{Schematics of a birth event}
\label{fig:appbirth}
	\end{figure}
The birth process involves a single parent particle at $({\bf x'}, \theta')$ producing an offspring at state $({\bf x}, \theta)$, with rate per particle $b G^B_{\bx' \bx \theta' \theta}(\bn)$ (see Fig.\,\ref{fig:appbirth}). The parent remains unchanged. The net effect is an increase of one individual in the population at $({\bf x}, \theta)$. The total rate for birth events that add one particle at $({\bf x}, \theta)$ is:
\begin{equation}
	W_B(\bn \rightarrow \bn + \bm{e}_{{\bf x}}^{\theta}) = b\;\sum_{{\bf x'},\theta'}^{} G^B_{\bx' \bx \theta' \theta}(\bn) \, n_{\bf x'}^{\theta'} \,,
	\label{app:birth_transition_rate}
\end{equation}
where $\bm{e}_{{\bf x}}^{\theta}$ is the unit vector for the state $({\bf x}, \theta)$. The sum and the occupation number factor indicates that the birth of the particle at $(\bx,\theta)$ can be originated from any particle at any state $(\bx',\theta')$. 

The master equation for a system in which only the birth process occurs will be:
\begin{align}
	\label{eq:birth_master_eq}
 &\frac{\partial P(\bn, t)}{\partial t} = b\sum_{\substack{\bx \theta}} 
\Bigg[ W_B(\bn- \bm{e}_\bx^\theta \rightarrow \bn) P(\bn- \bm{e}_\bx^\theta, t)  \nonumber\\ & 
 	- W_B(\bn \rightarrow \bn + \bm{e}_\bx^\theta) P(\bn, t) \Bigg] = \\
 & b\sum_{\substack{{\bf x,x}' \\ \theta,\theta'}} 
 \Bigg[  G^B_{\bx' \bx \theta'\theta}(\bn-\bm{e}_{{\bf x}}^{\theta} )\left(  n_{\bf x'}^{\theta'} -\delta_{{\bf x  x}'}\delta_{\theta \theta'} \right) P(\bn- \bm{e}_{{\bf x}}^{\theta}, t)  \nonumber\\ & 
 	- G^B_{\bx' \bx \theta'\theta}(\bn) n_{\bf x'}^{\theta'} \; P(\bn, t) \Bigg].
\end{align}

The term $\delta_{{\bf x  x}'}\delta_{\theta \theta'}$ takes into account that when $\bx=\bx'$ and $\theta=\theta'$, the number of particles in this state for this gain term is $n_{\bx'}^{\theta'} -1$. To derive equations for the average occupation numbers $\langle n_{\bf y}^\beta \rangle$, we take the time-derivative  of the expectation value as
\begin{equation}
	\frac{d}{d t}\langle n_{\bf y}^\beta \rangle = \frac{d}{d t} \sum_{\bn} n_{\bf y}^\beta P(\bn, t) = \sum_{\bn} n_{\bf y}^\beta \frac{\partial P(\bn, t)}{\partial t},
	\label{eq:birth_expectation_start}
\end{equation}
and from Eq.\,\eqref{eq:birth_master_eq}:
\begin{align}
	&\frac{d \langle n_{\bf y}^\beta \rangle}{d t} = b\; \sum_{\bn}
	\sum_{\substack{{\bf x,x'} \\ \theta,\theta'}} 
\times \nonumber\\ & \bigg\lbrack G^B_{\bx' \bx \theta'\theta}(\bn -\bm{e}_{{\bf x}}^{\theta}) n_{\bf y}^\beta ( n_{\bf x'}^{\theta'} -\delta_{{\bf x x}'}\delta_{\theta \theta'}) P(\bn - \bm{e}_{{\bf x}}^{\theta}, t) 
	\nonumber\\
	&-G^B_{\bx' \bx \theta'\theta}(\bn)n_{\bf y}^\beta n_{\bf x'}^{\theta'} P(\bn, t) \bigg\rbrack.
	\label{eq:birth_expectation_expanded}
\end{align}
The two terms of Eq.~(\ref{eq:birth_expectation_expanded}) can be treated separately. For the \emph{loss term} ($T^B_{\text{loss}}$):
\begin{align*}
	T^B_{\text{loss}} &= - b\;\sum_{\substack{{\bf x,x}' \\ \theta,\theta'}} \sum_{\bn}G^B_{\bx' \bx \theta'\theta}(\bn)  n_{\bf y}^\beta n_{\bf x'}^{\theta'} P(\bn, t) \\
	&= - b\;\sum_{\substack{{\bf x,x}' \\ \theta,\theta'}} \bigg\langle G^B_{\bx' \bx \theta'\theta}(\bn)  n_{\bf y}^\beta n_{\bf x'}^{\theta'} \bigg\rangle, 
\end{align*}
where the average is with the probability $P(\bn,t)$. For the \emph{gain term} ($T^B_{\text{gain}}$), we shift the summation variable $\bM = \bn - \bm{e}_{{\bf x}}^{\theta}$ (so $\bn = \bM + \bm{e}_{{\bf x}}^{\theta}$ and $n_{\bf y}^\beta = m_{\bf y}^\beta + \delta_{{\bf y x}}\delta_{\beta \theta}$):

\begin{align*}
	T^B_{\text{gain}} 
	&=    b\;\sum_{\substack{{\bf x,x}' \\ \theta,\theta'}} \sum_{\bM} G^B_{\bx' \bx \theta'\theta}(\bM)(m_{\bf y}^\beta + \delta_{{\bf y x}}\delta_{\beta \theta}) m_{\bf x'}^{\theta'}  P(\bM, t) \\
	&=    b\;\sum_{\substack{{\bf x,x}' \\ \theta,\theta'}}  \left[ \langle G^B_{\bx' \bx \theta'\theta}(\bn)n_{\bf y}^\beta n_{\bf x'}^{\theta'} \rangle + \delta_{{\bf y x}}\delta_{\beta \theta} \langle  G^B_{\bx' \bx \theta'\theta}(\bn) n_{\bf x'}^{\theta'} \rangle  \right]
\end{align*}

In the last expression we have used that $\bM$ is a dummy summation variable, so that we can change $\bM\rightarrow \bn$, and the last averages are then over $P(\bn,t)$. Combining $T^B_{\text{loss}}+T^B_{\text{gain}}$, the exact equation for the mean occupations is:
\begin{equation}
	\boxed{	\frac{d \langle n_{\bf y}^\beta \rangle}{d t} = b\, \sum_{\bx', \theta'}  \langle  G^B_{\bx' \by \theta'\beta}(\bn )n_{\bf x'}^{\theta'} \rangle\,.}
	\label{app:mean_birth}
\end{equation}

\subsection{Master Equation for Death Process}
\begin{figure}[h!]
	\includegraphics[width=0.95\columnwidth]{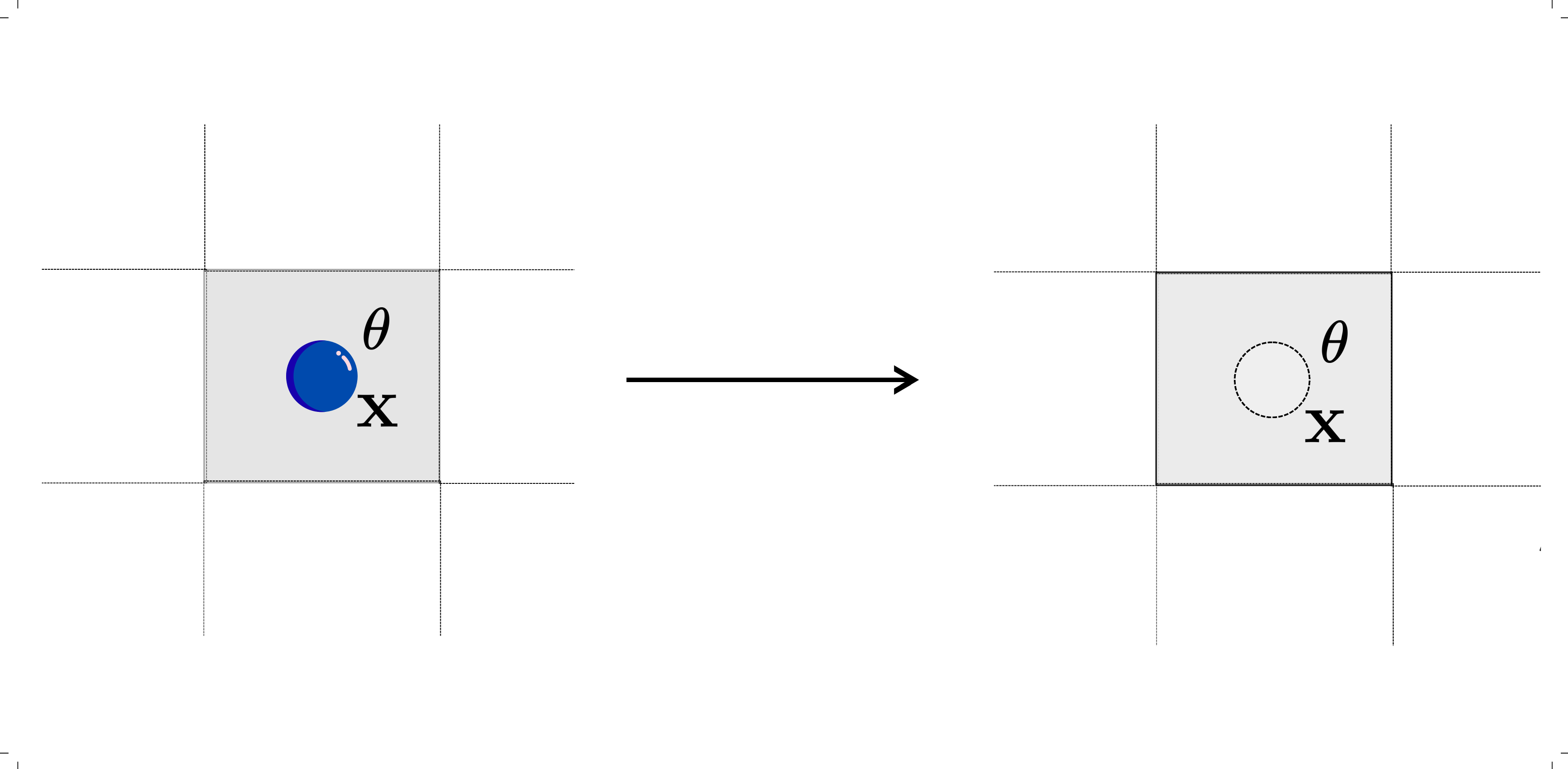}
	\label{fig:appdeath}
\caption{Schematics of a death event}
\end{figure}

Using that $d~G^D_{\bx,\theta}(\bn)$ is the probability rate per particle for death represented in Fig.\,\ref{fig:appdeath}, 
the total transition rate for the process that removes one particle precisely from $(\bx,\theta)$ is:
\begin{equation}
	W_D(\bn \rightarrow \bn - \bm{e}_{\bf x}^{\theta}) =  d~G^D_{\bx,\theta}(\bn)\,n_{\bf x}^\theta  \ .
	\label{app:death_transition_rate}
\end{equation}
The master equation for a system in which only the death process occurs, summing over all possible death locations:
\begin{align}
	\frac{\partial P(\bn, t)}{\partial t} &=  d \sum_{{\bf x},\theta} \bigg\lbrack G^D_{\bx,\theta}(\bn +\bm{e}_{\bf x}^{\theta}) (n_{\bf x}^\theta + 1) P(\bn + \bm{e}_{\bf x}^{\theta}, t) \nonumber\\
	&-G^D_{\bx,\theta}(\bn)\, n_{\bf x}^\theta  P(\bn, t) \bigg\rbrack \ .
	\label{eq:death_master_eq}
\end{align}
The average occupation evolves with:
\begin{align}
& \frac{d\langle n_{\bf y}^\beta\rangle}{dt} = \nonumber \\
& d \sum_{\bn,{\bf x},\theta}G^D_{\bx,\theta}(\bn+\bm{e}_{\bf x}^{\theta})n_{\bf y}^\beta \left[(n_{\bf x}^\theta + 1) P(\bn + \bm{e}_{\bf x}^{\theta}, t) - n_{\bf x}^\theta  P(\bn, t) \right].
\end{align}
The loss term is
\begin{align*}
	T^D_{\text{loss}} &= - d\sum_{{\bf x},\theta}\langle G^D_{\bx,\theta}(\bn) n_{\bf y}^\beta n_{\bf x}^\theta\rangle,
\end{align*}
and, doing again the shift $\bM=\bn + \bm{e}_{\bf x}^{\theta}$ and the renaming $\bM\rightarrow\bn$, the gain term is 
\begin{align*}
	T^D_{\text{gain}} &= 
	 d\sum_{{\bf x},\theta} \bigg\lbrack\langle G^D_{\bx,\theta}(\bn) n_{\bf y}^\beta n_{\bf x}^\theta \rangle - \langle G^D_{\bx,\theta}(\bn)\delta_{\bf xy}\delta_{\beta\theta} n_{\bf x}^\theta  \rangle \bigg\rbrack .
\end{align*}

Both terms combined give 
\begin{equation}
	\boxed{	\frac{d\langle n_{\bf y}^\beta\rangle}{dt}=  -d\langle G^D_{\by,\beta}(\bn)  n^\beta_{\bf y}\rangle} \label{app:mean_death}
\end{equation}

\subsection{Master Equation for Competition}
\begin{figure}[h!]
	\includegraphics[width=\columnwidth]{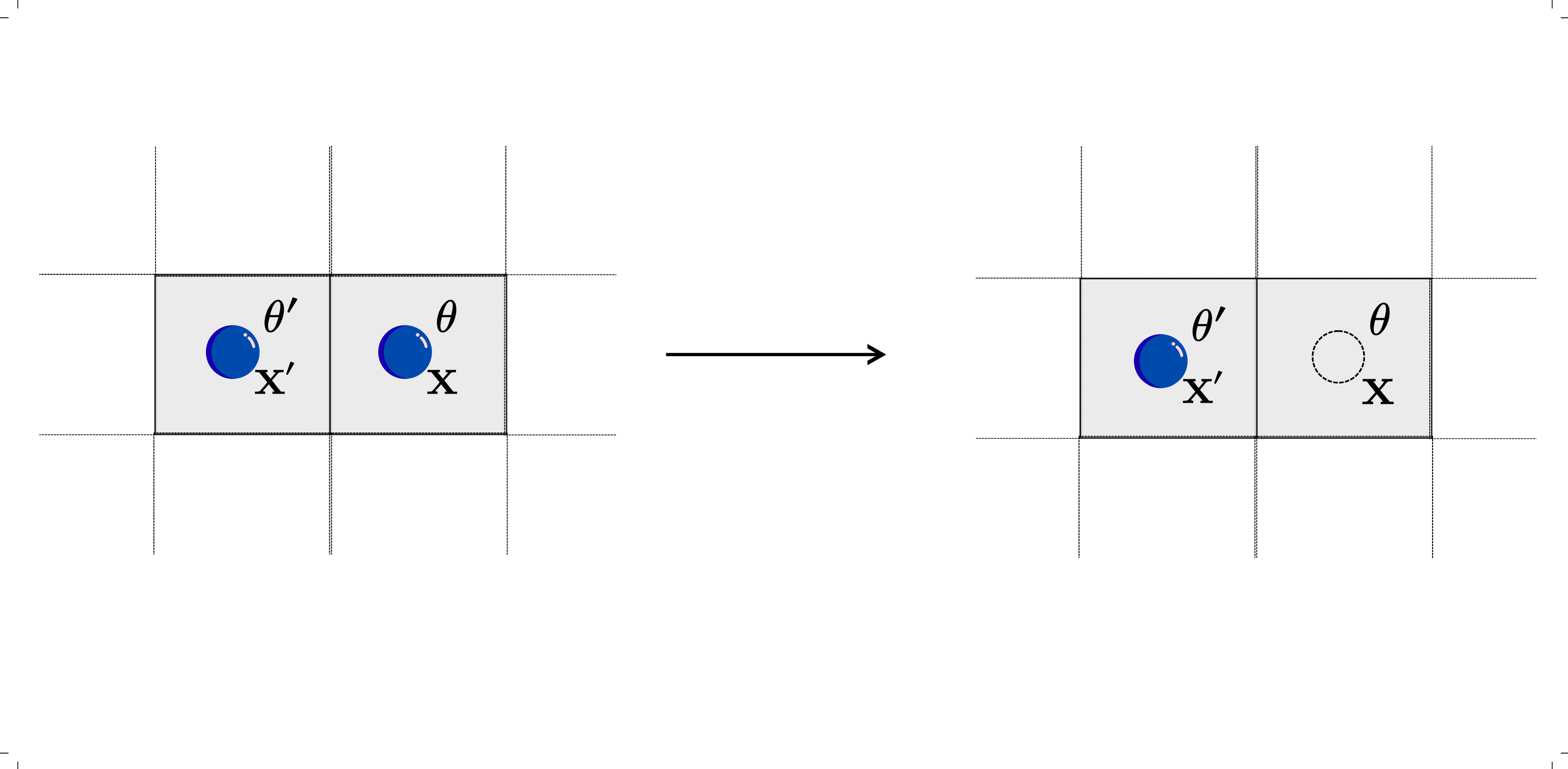}
	\label{fig:appcomp}
\caption{Schematics of the elimination of a particle by a competitive process}
\end{figure}
The transition rate per pair of particles for removing a particle from $(\bx,\theta)$ by its competitive interaction with a particle at $(\bx',\theta')$ is denoted by $\gamma G^C_{\bx'\bx\theta'\theta}(\bn)$ and represented in Fig.\,\ref{fig:appcomp}. The total transition rate for the process that removes one particle of type $({\bf x}, \theta)$ is given by the sum over all possible pairwise interactions with it:
\begin{equation}
	W_C(\bn \rightarrow \bn - \bm{e}_{{\bf x}}^{\theta}) =  \gamma\sum_{{\bf x'},\theta'}  G^C_{\bx'\bx\theta'\theta}(\bn)\, n_{\bf x}^\theta n_{\bf x'}^{\theta'} \ .
	\label{app:comp_transition_rate}
\end{equation}
To avoid unnecessary notational complications, we assume that a particle can compete with itself, so that $n_{\bf x}^\theta n_{\bf x'}^{\theta'}$ is the number of pair interacions even when $\bx=\bx'$ and $\theta=\theta'$. 
By summing over the states of all possible disappearing particles, the master equation for a system in which only this competitive process occurs is

\begin{align}
	&\frac{\partial  P(\bn, t)}{\partial t}=  \gamma\;\sum_{{\bf x,x}' ; \theta, \theta'} \bigg\lbrack G^C_{\bx'\bx\theta'\theta}(\bn +\bm{e}_{\bf x}^{\theta} )\times 
\notag	\\
	&\qquad(n_{\bf x}^\theta +1)(n_{\bf x'}^{\theta'} + \delta_{\bf xx'}\delta_{\theta\theta'})  P(\bn + \bm{e}_{\bf x}^{\theta}, t) \nonumber 
	\\
	&\qquad-G^C_{\bx'\bx\theta'\theta}(\bn) n_{\bf x}^\theta n_{\bf x'}^{\theta'} P(\bn, t) \bigg\rbrack
	\label{eq:master_eq}
\end{align}
As for the previous cases, we consider the average occupation numbers, shift the summation and combine the gain and loss terms to obtain 
\begin{equation}
	\boxed{	\frac{d\langle n_{\bf y}^\beta\rangle}{dt}= -\gamma \langle n^\beta_{\bf y}\sum_{{\bf x'},\theta'} G^C_{\bx'\by\theta'\beta}(\bn) n_{\bf x'}^{\theta'} \rangle} ,  \label{app:mean_comp}
\end{equation}
being the average over $P(\bn,t)$. 

\subsubsection*{Nonlinear birth and death as competitive interactions}
Here it is important to discuss the treatment of the averages in Eqs.\,\eqref{app:mean_birth}, \eqref{app:mean_death} and \eqref{app:mean_comp}. They are dependent on the possible dependence of the $G(\bn)$ kernels on the state, $\bn$. One way to handle those terms is to expand the kernels in a Taylor series on the system state. At linear order this can be written symbolically as 
\begin{align}
	G(\bn) = G^0 + \bn\cdot {\bf G}^1 + \cdots  \ ,
\label{eq:lindep}
\end{align}
where we have omitted the superindices $B$, $D$ and $C$, the subindices $\bx$, $\theta$, etc. $G^1$ is a vector with the same number of elements as $\bn=\{n_\bx^\theta\}$. 
 
To be specific, in the test model we used in the main text \cite{ehgclopezpre,v3gm-vhry}, the authors assume that death is independent of state $\bn$ and of location $(\bx,\theta)$, so that defining properly the death time-scale $d$ we can set $G^D_{\bx \theta} = 1$. Explicit pairwise competition is absent: $G^C_{\bx'\by \theta'\beta} = 0$. The microscopic birth process creates a new particle in the same state as the parent, so that $G^B_{\bx'\by\theta'\beta}\propto \delta_{\bx'\by}\delta_{\theta'\beta}$. Finally, the birth rate depends linearly (and negatively) on the occupations of some neighborhood of the birth location $(\bx,\theta)$. Thus, for this type of model, Eq. (\ref{eq:lindep}) reads: 
\begin{align}
	G^B_{\bx'\by \theta'\beta}(\bn) = \delta_{\bx'\by}\delta_{\theta'\beta} \left(1 - \sum_{\bz\delta} n_\bz^{\delta} F_{\bz\by\delta\beta} \right)\, .
\label{app:linearbirth}
\end{align}
The birth time-scale $b$ has been defined so that the state-independent term in this last equation is 1. 

Thus, summing up the demographic contributions to the change in the mean occupations: 
\begin{align}
\frac{d\langle n_\by^\beta \rangle}{dt} &= \eqref{app:mean_birth}+ \eqref{app:mean_death} + \eqref{app:mean_comp} = \nonumber \\
&(b - d)\langle n_\by^\beta \rangle - b\langle n_\by^\beta \sum_{\bx',\theta'} F_{\bx'\by \theta' \beta} n_{\bx'}^{\theta'} \rangle 
\label{app:sum_mean}
\end{align}

Comparing to Eq. (\ref{app:mean_comp}) we see that, for this type of model and for computing the evolution of average occupation numbers, a birth rate linearly dependent on the occupation numbers, as in (\ref{eq:lindep}), is equivalent to the combination of constant birth rate and a state-independent pairwise competition rate ($G^C(\bn)=F$). For state-dependent competition rates the correspondence would still hold after a mean-field approximation in which all factors in (\ref{app:mean_comp}) and in (\ref{app:sum_mean}) become replaced by their mean values (with respect to $P(\bn,t)$). It should be said that, since $G^B$ is a probability rate, it should remain positive. In numerical implementations of the particle dynamics this is enforced by replacing Eq. (\ref{app:linearbirth}) by zero when it becomes negative \cite{ehgclopezpre,v3gm-vhry}. We are here interested in developing a continuous-density description for which these positivity requirements do not apply directly. Nevertheless, in order to obtain a good description of the particle model, we will take care in the implementation of our explicit examples that quantities such as Eq. (\ref{app:linearbirth}) remain positive (using $b \gg d$). 

We can also consider a more general modelling framework, in which also the death rate depends linearly on the occupation of some neighbor sites \cite{ehgclopezpre,Birch2006}:
\begin{align}
d~G^D_{\by\beta} = d~\left( 1 + \sum_{\bz\delta} n_\bz^\delta H_{\bz\by\delta\beta}  \right) \ .
\label{app:lineardeath}
\end{align}
By substituting in Eq. (\ref{app:mean_death}) we see that a linear dependence of the death rate on occupations $\bn$ is equivalent to the sum of a occupation-independent death rate and an extra term with the structure of a competition interaction. 

In summary,  birth and death rates with linear dependence on the occupation of some neighborhood of the reproducing or dying individual, of the form (\ref{app:linearbirth}) and (\ref{app:lineardeath}), are equivalent to a combination of state-independent birth and death rates and competitive  interactions. More specifically, an explicit competition interaction $\gamma G^C$ gets replaced by $\gamma ~G^C + b~F + d~H$. This gives a well defined (\textit{i.e.} positive) competition rate as far as $F$ and $H$ remain positive, so that the effect of linear interactions is to decrease the birth rate (via (\ref{app:linearbirth})) or to increase the death rate (via (\ref{app:lineardeath})).

%

\subsection{Master Equation for Hopping Process}
\label{sec:2dhopping}

\begin{figure}[h!]
	\includegraphics[width=\columnwidth]{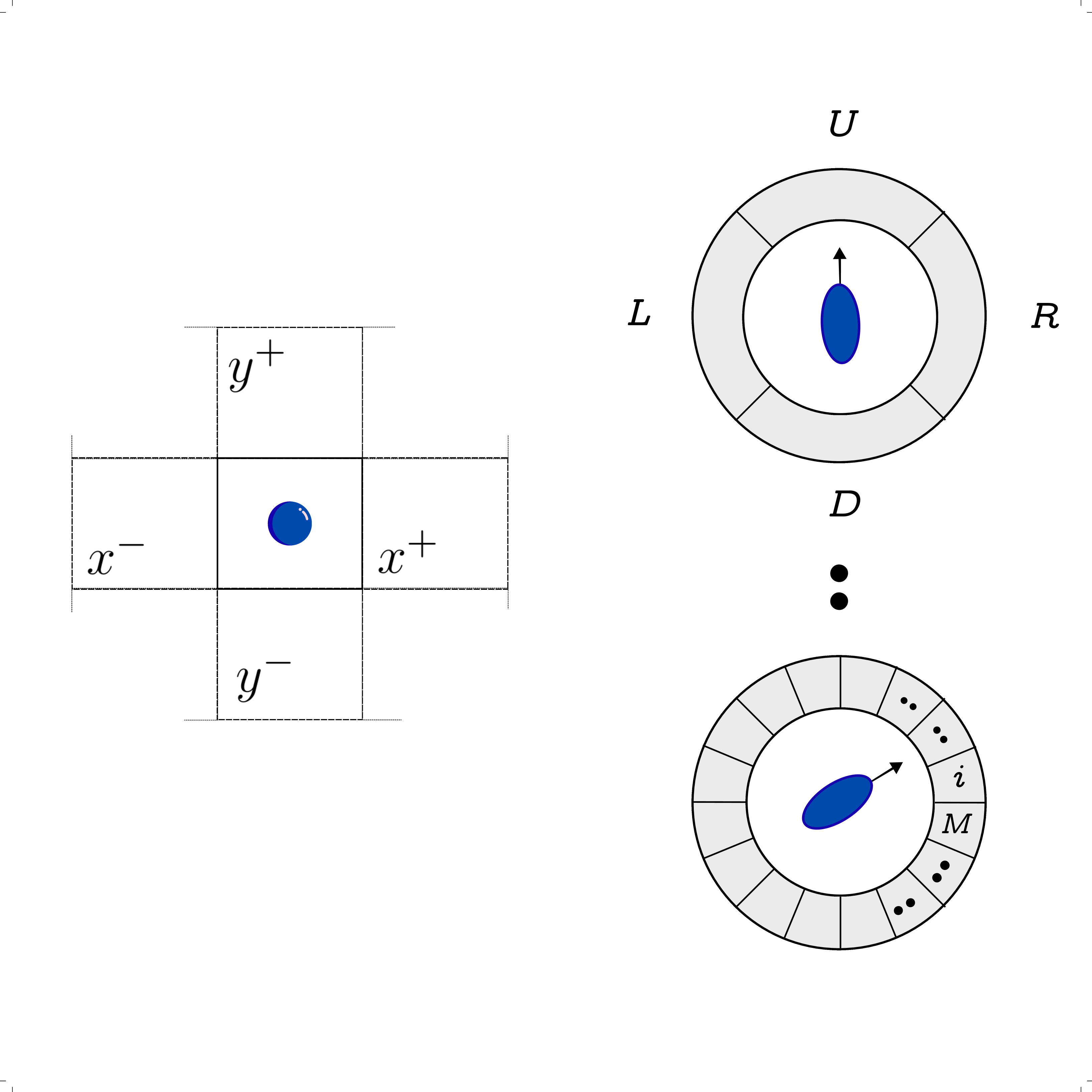}
	\caption{Left: Schematics of the 2D quadratic lattice grid and the hop directions of a particle. Right: the schematics of the orientational grid of a particle from the trivial 4 possible states to the $M$-finite discretization.} 
	\label{fig:app4}
\end{figure}

We denote the hopping rates per particle, from a location $(\bx,\theta)$, by $h^{\alpha}_{\bx,\theta}$, where $\alpha=U,D,R,L$, correspond to hopping along the four directions of the square lattice, $\{y^+,y^-,x^+,x^-\}$ respectively (see fig.\,\ref{fig:app4}). We restrict jumps to have the size of one lattice step, and for simplicity we will not consider any dependence of the hopping rates on the occupation numbers $\bn$. Due to the fact that our particles are active, some coupling is necessary between orientation and moving direction, \textit{i.e.} the dependence of the rates on the orientation angle $\theta$ is an important ingredient. In this section we just indicate this dependence in the notation. Later, when implementing a continuum limit in Sect. \ref{sec:continuum_MF} we will specify an explicit dependence leading to a standard continuous active-particle phenomenology. 

In terms of the hopping rates per particle, the system hopping rates are
\begin{align}
	W(\bn \rightarrow \bn - \bm{e}_{\mathbf{x}}^{\theta} + \bm{e}_{\mathbf{x}+\boldsymbol{\varepsilon}_1}^{\theta}) &= h^{R}_{\mathbf{x},\theta}\;n_{\mathbf{x}}^\theta 
\label{eq:2d_compact_hopping_transition_rate_R}\\
	W(\bn \rightarrow \bn - \bm{e}_{\mathbf{x}}^{\theta} + \bm{e}_{\mathbf{x}-\boldsymbol{\varepsilon}_1}^{\theta}) &= h^{L}_{\mathbf{x},\theta}\;n_{\mathbf{x}}^\theta 
\label{eq:2d_compact_hopping_transition_rate_L}\\
	W(\bn \rightarrow \bn - \bm{e}_{\mathbf{x}}^{\theta} + \bm{e}_{\mathbf{x}+\boldsymbol{\varepsilon}_2}^{\theta}) &= h^{U}_{\mathbf{x},\theta}\;n_{\mathbf{x}}^\theta 
	\label{eq:2d_compact_hopping_transition_rate_U}\\
	W(\bn \rightarrow \bn - \bm{e}_{\mathbf{x}}^{\theta} + \bm{e}_{\mathbf{x}-\boldsymbol{\varepsilon}_2}^{\theta}) &= h^{D}_{\mathbf{x},\theta}\;n_{\mathbf{x}}^\theta
	\label{eq:2d_compact_hopping_transition_rate_D}
\end{align}
where $\boldsymbol{\varepsilon}_1 = (1,0)$ and $\boldsymbol{\varepsilon}_2 = (0,1)$ are basis vectors on the square lattice.

The master equation implementing these hopping transitions can be written compactly as:
\begin{align}
	&\frac{\partial P(\bn, t)}{\partial t} = \sum_{\mathbf{x},\theta} \sum_{\alpha\in\{L,R,U,D\}} \times \nonumber\\
	&\Bigg[ 
	h^{\alpha}_{\mathbf{x}-\boldsymbol{\varepsilon}_\alpha,\theta} \big(n_{\mathbf{x}-\boldsymbol{\varepsilon}_\alpha}^\theta + 1\big) P(\bn + \bm{e}_{\mathbf{x}-\boldsymbol{\varepsilon}_\alpha}^{\theta} - \bm{e}_{\mathbf{x}}^{\theta}, t) 
	- h^{\alpha}_{\mathbf{x},\theta} n_{\mathbf{x}}^\theta P(\bn, t) \Bigg],
	\label{eq:2d_compact_hopping_master_eq}
\end{align}
where $\boldsymbol{\varepsilon}_R = \boldsymbol{\varepsilon}_1$, $\boldsymbol{\varepsilon}_L = -\boldsymbol{\varepsilon}_1$, $\boldsymbol{\varepsilon}_U = \boldsymbol{\varepsilon}_2$, $\boldsymbol{\varepsilon}_D = -\boldsymbol{\varepsilon}_2$.
The average occupation numbers evolve as
\begin{align}
	\frac{d\langle n_{\mathbf{y}}^\beta\rangle}{dt} &= \sum_{\bn} n_{\mathbf{y}}^\beta \sum_{\mathbf{x},\theta} \sum_{\alpha} \times\nonumber\\
	&\Bigg[ 
	h^{\alpha}_{\mathbf{x}-\boldsymbol{\varepsilon}_\alpha,\theta} \big(n_{\mathbf{x}-\boldsymbol{\varepsilon}_\alpha}^\theta + 1\big) P(\bn + \bm{e}_{\mathbf{x}-\boldsymbol{\varepsilon}_\alpha}^{\theta} - \bm{e}_{\mathbf{x}}^{\theta}, t) - h^{\alpha}_{\mathbf{x},\theta} n_{\mathbf{x}}^\theta P(\bn t) \Bigg].
\end{align}
The loss term is
\begin{align*}
	T^H_{\text{loss}} = -\sum_{\mathbf{x},\theta} \sum_{\alpha} h^{\alpha}_{\mathbf{x},\theta} \langle n_{\mathbf{y}}^\beta n_{\mathbf{x}}^\theta \rangle , 
\end{align*}
and the gain terms are
\begin{align*}
	&T^H_{\text{gain}}= \sum_{\mathbf{x},\theta} \sum_{\alpha} \times 
	\\
	 &h^{\alpha}_{\mathbf{x}-\boldsymbol{\varepsilon}_\alpha,\theta} \bigg\langle \Big(n_{\mathbf{y}}^\beta + (\delta_{\mathbf{x},\mathbf{y}} - \delta_{\mathbf{x}-\boldsymbol{\varepsilon}_\alpha,\mathbf{y}})\delta_{\beta\theta}\Big) n_{\mathbf{x}-\boldsymbol{\varepsilon}_\alpha}^\theta \bigg\rangle \\
	&= \sum_{\mathbf{x},\theta} \sum_{\alpha}  \quad h^{\alpha}_{\mathbf{x}-\boldsymbol{\varepsilon}_\alpha,\theta} \langle n_{\mathbf{y}}^\beta n_{\mathbf{x}-\boldsymbol{\varepsilon}_\alpha}^\theta \rangle + \\ &\quad \sum_{\alpha} \bigg\lbrace   h^{\alpha}_{\mathbf{y}-\boldsymbol{\varepsilon}_\alpha,\beta} \langle n_{\mathbf{y}-\boldsymbol{\varepsilon}_\alpha}^\beta \rangle
	 -  h^{\alpha}_{\mathbf{y},\beta}\langle n^\beta_{\mathbf{y}} \rangle \bigg\rbrace,
\end{align*}


Combining both terms we get the compact form:
\begin{equation}
		\Scale[0.94]{
	\boxed{\frac{d\langle n_{\mathbf{y}}^\beta\rangle}{dt}= \sum_{\alpha\in\{L,R,U,D\}} \bigg\lbrack h^{\alpha}_{\mathbf{y}-\boldsymbol{\varepsilon}_\alpha,\beta} \langle n_{\mathbf{y}-\boldsymbol{\varepsilon}_\alpha}^\beta \rangle - h^{\alpha}_{\mathbf{y},\beta} \langle n_{\mathbf{y}}^\beta \rangle \bigg\rbrack  .}
}
\end{equation}


\subsection{Master Equation for Angular Tumbling}


The tumbling process consists of a particle at state $({\bf x},\theta)$ changing its internal orientation while remaining at the same spatial position ${\bf x}$. For 2D the easiest assumption is that the orientation has the same number of states as the directions of motion in the lattice, \textit{i.e.} $\theta =\{U,D,R,L\}$.  
However, this will not generate the equations of motion for a particle that has a free continuous two-dimensional orientation. We must increase the number of states in the variable $\theta =\{1,\cdots , M\}$ such that in the limit $M\rightarrow \infty$ we have a continuous angle. We restrict to jumps to a neighboring orientation: $\theta\rightarrow \theta+1$ and $\theta\rightarrow \theta-1$ with rates per particle $t^+_{\bx\theta}$ and $t^-_{\bx\theta}$, respectively (with periodic boundary conditions, so that $M+1=1$). We assume that tumbling rates are independent of occupation numbers.  
The system transition rates become:
\begin{align}
	W^+_T(\bn \rightarrow \bn - \bm{e}_{\bf x}^{\theta} + \bm{e}_{\bf x}^{\theta+1}) &=  t^+_{\bx\theta}\,n_{\bf x}^\theta \;\quad  \\
	W^-_T(\bn \rightarrow \bn - \bm{e}_{\bf x}^{\theta} + \bm{e}_{\bf x}^{\theta-1}) &=  t^-_{\bx\theta} \,n_{\bf x}^\theta \;.\quad 
	\label{eq:state_dep_tumbling_transition_rate}
\end{align}

The corresponding master equation for a system in which only tumbling acts reads:
\begin{align}
	\frac{\partial P(\bn, t)}{\partial t} &= 
	\sum_{{\bf x},\theta} \bigg\lbrack 
	t^+_{\bx\theta-1} \,(n_{\bf x}^{\theta-1}+1)\,
	P(\bn + \bm{e}_{\bf x}^{\theta-1} - \bm{e}_{\bf x}^{\theta}, t)
\nonumber\\
	&\quad + t^-_{\bx\theta+1} \,(n_{\bf x}^{\theta+1}+1)\,
	P(\bn + \bm{e}_{\bf x}^{\theta+1} - \bm{e}_{\bf x}^{\theta}, t)
\nonumber\\
&\quad	- (t^+_{\bx\theta} + t^-_{\bx\theta}) \, n_{\bf x}^{\theta} P(\bn, t)
	\bigg\rbrack .
	\label{eq:tumbling_master_eq}
\end{align}
The time evolution of the occupation numbers is given by 
\begin{align}
	\frac{d\langle n_{\bf y}^\beta\rangle}{dt} &=
	\sum_{\bn} n_{\bf y}^\beta 
	\sum_{{\bf x},\theta} \bigg\lbrack 
	t^+_{\bx\theta-1} \,(n_{\bf x}^{\theta-1}+1)\,
	P(\bn + \bm{e}_{\bf x}^{\theta-1} - \bm{e}_{\bf x}^{\theta}, t)
\nonumber\\
	&\quad + t^-_{\bx\theta+1} \,(n_{\bf x}^{\theta+1}+1)\,
	P(\bn + \bm{e}_{\bf x}^{\theta+1} - \bm{e}_{\bf x}^{\theta}, t)
\nonumber\\
&\quad	- (t^+_{\bx \theta} + t^-_{\bx \theta})\, n_{\bf x}^{\theta} P(\bn, t)
	\bigg\rbrack .
\end{align}

%

Manipulating this expression as before, we obtain the final expression for the evolution of the average occupations due to tumbling: 
\begin{equation}
	\Scale[0.88]{
		\boxed{
			\frac{d\langle n_{\bf y}^\beta\rangle}{dt}
			=
			t^+_{\by \beta-1}\langle n_{\bf y}^{\beta-1}\rangle
			+
			t^-_{\by \beta+1}\langle n_{\bf y}^{\beta+1}\rangle
			-
			(t^+_{\by \beta} + t^-_{\by \beta})\langle n_{\bf y}^\beta\rangle
	}
}
	\;.
\label{app:tumbling}
\end{equation}

	\section{Mean field and continuum limit}\label{sec:continuum_MF}

	To set up a continuous approximation for the average occupations of previous section we need to introduce the expected density $\rho(\by,\beta,t)$ by considering the angular spacing between the $M$ allowed interactions, $\Delta\beta$, and the spatial displacement vectors associated to the distances between the grid sites,  $\boldsymbol{\varepsilon}_\alpha \Delta y$, with modulus $\Delta y$ (we assume isotropy). The elemental  grid volume is $(\Delta y)^d$ (typically, here $d=2$). {Note that we find more convenient to use in this Appendix the symbols $\Delta\beta$ and $\Delta y$ for the angula and spatial grid spacings. They are the same quantities that were named $\Delta \theta$ and $\Delta x$ in the main text.} In terms of these quantities we introduce the following replacements:  
\begin{align}
\label{app:startsubs}
       &  \boldsymbol{\varepsilon}_\alpha \rightarrow  \boldsymbol{\varepsilon}_\alpha \Delta y \\
		&\langle n_{\bf y}^\beta (t)\rangle \rightarrow \Delta \beta \Delta y^d\, \rho({\bf y},\beta,t)
	\\
	& \langle n_{\by + \boldsymbol{\varepsilon}_\alpha}^{\beta\pm 1} (t) \rangle\rightarrow \Delta \beta \Delta y^d \, \rho({\bf y} + \boldsymbol{\varepsilon}_\alpha \Delta {\bf y},\beta\pm\Delta \beta,t) 
	\\
	& \sum_{\by\beta} \rightarrow \frac{1}{\Delta \beta (\Delta y)^d} \int d\by d\beta  
	\\
    & G^B_{\bx\by\theta\beta}(\bn) \rightarrow G^B(\bx,\by,\theta,\beta)(\Delta y)^d \Delta\beta 
\\
   & G^D_{\by \beta}(\bn) \rightarrow G^D(\by,\beta)
   \\
   & G^C_{\bx\by\theta\beta}(\bn) \rightarrow G^C(\bx,\by,\theta,\beta)
\\
	& h^{\alpha}_{{\bf y}+\boldsymbol{\varepsilon}_\alpha,\beta} =  h^{\alpha}({\bf y}+\boldsymbol{\varepsilon}_\alpha \Delta y,\beta) 
\\
   & t^\pm_{\by\beta\pm 1} \rightarrow t^\pm(\by,\beta\pm \Delta\beta) \ ,
   \label{app:endsubs}
\end{align}
and take the limit $\Delta\beta, ~\Delta y\rightarrow 0$. For the death and competition kernels, $G^D$ and $G^C$, the approximation is just a change of notation. For the birth kernel the above renormalization is needed to achieve an adequate continuous limit.

	For simplicity, in writing the above substitutions for the rates, we have assumed that they depend only on the spatial and angular indices, not in the full state $\bn$ of the system. For the demographic rates (birth, death and competition) there is no problem (after implementing the mean-field approximation) in including such dependence. In addition, as shown before, for the case of birth and death kernels with linear dependence on this state, the dependence can be incorporated as a modified competition rate. To further obtain a closed evolution equation for $\rho({\bf y},\beta,t)$ we make the mean-field approximation, in which averages of products factorize, e.g. in Eq. (\ref{app:mean_comp}) $\langle n_\by^\beta n_{\bx'}^{\theta'}\rangle \approx  \langle n_\by^\beta\rangle \langle n_{\bx'}^{\theta'}\rangle$. 

\subsection{Birth, death, competition}

By introducing the substitutions (\ref{app:startsubs})-(\ref{app:endsubs}) in the sum of Eqs. (\ref{app:mean_birth}), (\ref{app:mean_death}) and (\ref{app:mean_comp}), and performing the mean field approximation and the limits, it is straightforward to obtain 
\begin{align}
	&\frac{d\rho(\by,\beta,t)}{dt} = b \int d{\bf x'} d\theta'\;  G^B(\bx', \by,\theta',\beta)  \rho({\bf x'},\theta',t)  \nonumber 
	\\
	& - d ~G^D(\by,\beta)\rho({\bf y},\beta,t) \nonumber 
   \\
	& - \gamma \,\rho({\bf y},\beta,t)\int d{\bf x'} d\theta' \; G^C(\bx', \by,\theta',\beta)  \rho({\bf x'},\theta',t) \ , 
 \label{app:BDC_meanfield}
\end{align}
for processes involving only the demographic processes (birth, death and  competition). {These are the demographic terms that are included in Eq.\eqref{eq:final_result} (after the trivial notation replacement $\by,\beta$ $\rightarrow$ $\bx,\theta$}.

	\subsection{Two-dimensional hopping}

	The expansion at second order of the hopping contribution in 2D can be written in a compact form as,
	\begin{align}
\frac{d\rho(\by,\beta,t)}{dt} &= - \boldsymbol{\nabla} \cdot \bigg\lbrack \rho({\by},\beta,t) \sum_\alpha \boldsymbol{\varepsilon}_\alpha h^\alpha({\bf y},\beta,t)\bigg\rbrack  \Delta y
	\nonumber	\\
 & + \frac{(\Delta y)^2}{2} \nabla^2 \bigg\lbrack \rho(\by,\beta,t) \sum_\alpha h^\alpha(\by,\beta) \bigg\rbrack  \ . 
 \label{app:hop2d}
	\end{align}
%
We now specify a set of rates that in the continuum limit would give rise to a standard model of self-propulsion \cite{digregorio}, in which the spatial motion of particles occurs in the direction of its orientation angle. We have only four directions of motion in the lattice, $U$, $D$, $L$ and $R$, and $M\rightarrow \infty$ orientations. A natural way to specify motion along the four lattice directions is to weight the corresponding jumping rates according to the projection of the orientation direction along the lattice direction, namely 
	 	\begin{align}
\label{app:startjumprates}
	 	&h^U({\bf y},\beta) = \frac{D_T}{2 (\Delta y)^2} + \frac{v_0}{2\Delta y} \sin (\beta)
	 	\\
	 	&h^D({\bf y},\beta) = \frac{D_T}{2 (\Delta y)^2} - \frac{v_0}{2\Delta y} \sin(\beta)
	 	\\
	 	&h^R({\bf y},\beta) = \frac{D_T}{2 (\Delta y)^2} + \frac{v_0}{2\Delta y} \cos(\beta)
	 	\\
	 	&h^L({\bf y},\beta) =\frac{D_T}{2 (\Delta y)^2} - \frac{v_0}{2\Delta y} \cos(\beta) \ .
\label{app:endjumprates}
	 \end{align}
$v_0$ and $D_T$ are constants independent of $\by$ and $\beta$, so that these jumping rates only depend on orientation $\beta$, not position $\by$: $h^\alpha(\by,\beta)=h^\alpha(\beta)$. In the continuous limit, $\Delta y\rightarrow 0$, (\ref{app:hop2d}) becomes
	 \begin{align}
\frac{d\rho(\by,\beta,t)}{dt} = - v_0 \boldsymbol{\nabla}\cdot (\hat{\mathbf{n}}(\beta) \rho({\bf y}, \beta, t)) + D_T \boldsymbol{\nabla}^2 \rho({\bf y}, \beta, t), \label{app:final_hop2d}
	 \end{align}
	 with $\hat{\mathbf{n}}(\beta)  = (\cos(\beta), \sin(\beta))$, the orientation unit vector. This equation shows that the parameters $v_0$ and $D_T$ entering in the expressions (\ref{app:startjumprates})-(\ref{app:endjumprates}) have the meaning of the modulus of a self-propulsion velocity and a translational diffusion coefficient, respectively. 
	 
\subsection{Tumbling}
Expansion of Eq.\,(\ref{app:tumbling}) for small $\Delta \beta$ gives 
\begin{align}
	&\frac{d\rho(\by,\beta,t)}{dt} = \Delta \beta \;\partial_\beta \bigg\lbrack \rho({\bf y},\beta,t) \bigg(t^-(\by,\beta)- t^+(\by,\beta)\bigg)\bigg\rbrack
	\nonumber\\
	&+\frac{(\Delta \beta)^2}{2}\partial^2_\beta \bigg\lbrack \rho({\bf y},\beta,t)\bigg(t^-(\by,\beta)+ t^+(\by,\beta)\bigg)\bigg\rbrack
\end{align}
	We assume now that the tumbling rates are symmetric ($t^+ = t^-$), and express them in terms of a rotational diffusion coefficient $t^+(\by,\beta)=t^-(\by,\beta) \equiv D_R(\by,\beta)/(\Delta\beta)^2$ so that the tumbling dynamics of the continuous density becomes
	\begin{align}
		\frac{d\rho(\by,\beta,t)}{dt} = \partial_\beta^2 (D_R(\by, \beta) \rho({\bf y},\beta,t))\,. \label{app:final_tumble}
	\end{align}

In the following, and in transfering this result to the complete expression in Eq. (\ref{eq:final_result}) we have made the additional simplifying assumption $D_R(\by, \beta) = D_R, \;\; \forall\; \by, \beta$.

\section{Angular-moments representation and closure}
	\label{app:momentum_exp}
\renewcommand{\theequation}{C\arabic{equation}}
\setcounter{equation}{0}  
\renewcommand{\thefigure}{C\arabic{figure}}
\setcounter{figure}{0}  

First, we perform the angular Fourier transform of each of the terms appearing in the mean-field equation for the mean density $\rho(\bx,\theta,t)$, Eq.\,\eqref{eq:final_result}. To do this we multiply each term by $\int e^{in\theta} d\theta$ and perform the integration. All angular integrals are over the interval $[0,2\pi]$.

For the spatial diffusion term, 
		\begin{align}
\int e^{in\theta} D_T\boldsymbol{\nabla}^2 \rho({\bf x},\theta,t)d\theta &= D_T\boldsymbol{\nabla}^2\int e^{in\theta} \rho({\bf x},\theta,t)d\theta \nonumber\\
		&= D_T\boldsymbol{\nabla}^2 \rho_n({\bf x},t)
	\end{align}
	
The angular diffusion term, 
	\begin{align}
\int e^{in\theta}& D_R\partial_\theta^2 \rho(\bx,\theta,t) d\theta= D_R \int \partial^2_\theta\left(e^{i n\theta}\right) \rho(\bx,\theta,t) d\theta \nonumber \\		&=  -n^2 D_R \rho_n(\bx,t)
	\end{align} 

To transform the selpropulsion contribution, we must use the identities:
	\begin{align}
		&\hat{\mathbf{n}}(\theta)\cdot\boldsymbol{\nabla} = \cos(\theta)\partial_{x} + \sin(\theta)\partial_y\,,
		\\
		&\cos(\theta) = \frac{1}{2}\{e^{i\theta} +e^{-i\theta}\}\,,
		\\
		&\sin(\theta) = \frac{1}{2 i}\{e^{i\theta} -e^{-i\theta}\} \ ,
	\end{align}
to show that,
	\begin{align}
		&- v_0 \int d\theta e^{in\theta} \hat{\mathbf{n}}(\theta)\cdot\boldsymbol{\nabla}\rho(\bx,\theta,t)	=\nonumber\\&= - \frac{v_0}{2}\bigg\lbrace (\partial_{ x} -i\partial_y)\rho_{n+1} +(\partial_{ x}+i\partial_y)\rho_{n-1}\bigg\rbrace\,.
	\end{align}
	
For the demographic terms (birth, death and competition) we recall the factorization, homogeneity and isotropy assumptions stated in the main text: $G^D(\bx,\theta)=1$, $G^{B,C}(\by,\bx,\beta,\theta)=S^{B,C}(|\by-\bx|)\Phi^{B,C}(|\beta-\theta|)$. 
With these assumptions, the death term does not need any special mathematical manipulation: 
	\begin{align}
		d\int e^{in\theta} \rho({\bf x},\theta,t) = d~ \rho_n({\bf x},t).
	\end{align} 
The angular Fourier transform of the birth term is
	\begin{align}
\Delta_n^B({\bf x},t)& =	\int d\theta e^{in\theta}~b \int d{\bf y} d\beta S^B(\bx-\by) \Phi^B(\theta-\beta) \rho({\bf y},\beta,t)
		\nonumber\\
&= b \int d\by S^B(\bx-\by) \sum_{m k} \rho_k(\by,t) \Phi^B_m \delta_{n m}\delta_{m k} \nonumber \\
&= b \Phi^B_n \int d\by S^B(\bx-\by) \rho_n(\by,t)
\nonumber
\\& \equiv b\Phi^B_n \left( S^B*\rho_n \right)(\bx,t) \ .
	\end{align}	
We have introduced both the inverse angular Fourier transform of $\rho$, Eq. (\ref{eq:rho_n_inverse}), and the one for $\Phi^B$:
\begin{align}
\Phi^B(\theta)=\frac{1}{2\pi}\sum_m \Phi^B_m e^{-i m \theta} \ ,
	\end{align}
in addition to (in the last line) the definition of the spatial convolution. 

	The same type of manipulations can be applied to the competition term, to obtain 
	\begin{align}
&\Delta_n^C(\bx,t) = \nonumber \\
-\gamma \int d\theta e^{i n\theta} & \rho(\bx,\theta) \int d{\bf y} d\beta S^C(\bx-\by) \Phi^C(\theta-\beta) \rho(\by,\beta,t)
	\nonumber	\\
&=- \frac{\gamma}{2\pi} \sum_m \rho_{n-m}(\bx,t) \left( S^C*\rho_m \right)(\bx,t) \Phi_m \ .
	\end{align}

%
%

The resulting set of equations for the evolution of angular modes $\rho_m(\bx,t)$ is (\ref{eq:rhon_modes}). The evolution equation for the $n=0$ mode, $\rho_0(\bx,t)=\rho(\bx,t)$, can be written exactly in terms of the density and polarization fields, $\rho$ and $\bP$, leading to Eq. (\ref{eq:rho_final}). 
The $n=1$ equation, however, involves $\rho_2$,
\begin{align}
	\partial_t \rho_1 &= \Delta_1^B+\Delta_1^C - (d+D_R)\rho_1 \nonumber\\&\qquad+ D_T\boldsymbol\nabla^2\rho_1
	 -\frac{v_0}{2}\Big[\partial_+\rho_0 + \partial_-\rho_2\Big],
	\\& \partial_\pm\equiv\partial_x\pm i\partial_y \ .
\end{align}
Similarly, the equation for $\rho_{-1}$ involves $\rho_{-2}$.
Under the bare first-order truncation $\rho_2$ is simply discarded ($\rho_{\pm 2}=0$). The
von Mises closure instead determines $\rho_2$ from 
Eq.\,\eqref{eq:vonmises_moments}. For $n=2$, and eliminating $e^{2i\phi}$ in
favour of $\rho_1$ via this same equation at $n=1$,
\begin{equation}
	\rho_2(\bx,t) = B\big(\kappa(\bx,t)\big)\, \frac{\rho_1(\bx,t)^2}{\rho(\bx,t)} ,
	\quad
B(\kappa)\equiv \frac{I_0(\kappa) I_2(\kappa)}{I_1(\kappa)^2} .
	\label{app:rho2_closure}
\end{equation}
$ B(\kappa)$ interpolates monotonically between $ B(\kappa) \rightarrow 1/2$ for $\kappa\to 0$ 
(weak polar order) and $ B(\kappa)\to 1$ as $\kappa\to\infty$
(strong polar order). The same computation can be performed for $n =-2$ giving.
\begin{align}
	\rho_{-2} =  B\big(\kappa(\bx,t)\big)\, \frac{\rho_{-1}(\bx,t)^2}{\rho(\bx,t)} . \label{app:closure2}
\end{align}

Carrying Eqs.~\eqref{app:rho2_closure},\eqref{app:closure2} through the equation for $\rho_{\pm 1}$ and 
introducing the polarization vector $\bP$ gives
\begin{align}
	\partial_t \bP = & D_T\nabla^2\bP - D_R\bP - d~\bP + \Delta_1^B + \Delta_1^C \nonumber \\
& -\frac{v_0}{2} \boldsymbol\nabla\rho
	- v_0\boldsymbol\nabla\cdot  {\bf Q}[\rho,\bP] \ ,
	\label{app:P_final}
\end{align}
which includes a finite quadrupolar (nematic-like) term containing the symmetric, traceless tensor
\begin{equation}
	{\bf Q}_{ij}(\bx,t) = \rho(\bx,t){\mathcal B(\kappa(\bx,t))}
	\Big(\frac{P_i(\bx,t)P_j(\bx,t)}{|\bP(\bx,t)|^2} - \frac{\delta_{ij}}{2}\Big) \ ,
	\label{app:T_tensor}
\end{equation}
where $i,j=x,y$ and 
\begin{align}
\mathcal B(\kappa) = {B(\kappa)}{\mathcal A^2(\kappa)} = \frac{I_2(\kappa)}{I_0(\kappa)}.
\end{align}
{At this point we summarize the hypothesis and approximations introduced above and bounding the validity of Eqs.\eqref{eq:rho_final} and \eqref{eq:P_final} of the main text. First, we have restricted our description to systems in which the birth and competition kernels factorize in a spatial and an orientational part. Second, we have imposed the kind of active motion specified by Eqs.(\ref{app:startjumprates})-(\ref{app:endjumprates}), which corresponds in the continuous limit to standard classes of active-particle dynamics. Third, we restrict to spatially and orientationally homogeneous and isotropic dynamics, from which parameters such as $D_T$, $D_R$ or $d$ become constants. Then, the continuous limit is performed and two approximations introduced: The mean field approximation, from which averages of the type $\langle n_y^\beta f(n)\rangle$ factorize as $\langle n_y^\beta\rangle\langle  f(n)\rangle$, and the von Mises closure Eq.\,\eqref{eq:vonmises_ansatz}, that allows to approximate all angular Fourier modes in terms of the lowest one. The mean-field approximation is expected to give reasonable results in locations where particle density $\rho(x,t)$ is not too small \cite{lawPOPULATIONGROWTHSPACE2003,morozovSpatiallyExplicitEcological2012}. On the other hand, a critical assessment of the von Mises closure approximation can be found in \cite{kurstenCriticalAssessmentMises2017}}.

In order to obtain a simpler model for easier numerical and analytical study, we have introduced a quasi-one-dimensional system, 
in which spatial motion is restricted to one direction (the horizontal $x$ one) but angular orientation is still free to rotate in 
two dimensions. Restriction to horizontal motion can be done by setting to zero the up and down jumping rates (B12) and (B13), from 
which the self-propulsion velocity is $v_0(\cos\theta,0)$, and Eq.\,(C3) becomes $\hat\bn(\theta)\cdot\nabla=\cos(\theta)\partial_x$. 
Repeating the derivations above, neglecting any dependence of the density and polarization fields on the $y$ direction, we find the 
following coupled equations for the polarization vector (only the self-propulsion terms, i.e. those containing $v_0$, are written):
\begin{align}
	\partial_t P_x &= -\frac{v_0}{2} \bigg\lbrack \partial_x \rho_0 + \partial_x\bigg( \frac{B(\kappa)}{\rho_0}(P_x^2 - P_y^2)\bigg) \bigg\rbrack\,,
	\\
		\partial_t P_y &= -\frac{v_0}{2} \bigg\lbrack \partial_x \bigg(2 \frac{B(\kappa)}{\rho_0} P_xP_y\bigg) \bigg\rbrack.
\end{align}
From these equations (even when combined with the remaining not shown density evolution terms) we see that if initially $\bP=(P_x,0)$, 
this horizontal polarization remains a solution at all times, so that the self-propulsion part of the polarization equation becomes
	\begin{align}
		\partial_t P_x = -\frac{v_0}{2}\partial_x \bigg( \rho_0 + \rho_0 \mathcal{B}(\kappa)\bigg)
	\end{align}
    \section{Histeresis and multistability}
\label{sec:app_hysteresis}
\renewcommand{\theequation}{D\arabic{equation}}
\setcounter{equation}{0}  
\renewcommand{\thefigure}{D\arabic{figure}}
\setcounter{figure}{0}  

The phase diagrams in Figs.\,\ref{fig:phase_diagram1d} and
\ref{fig:phase_diagram2d} of the main text describe the final states
reached from initial conditions consisting in the homogeneous solution
slightly perturbed by noise. We noticed that, both in one and in two
dimensions, other states coexist with those, being reached from other
initial conditions. To give a preliminary idea of this multistability we
describe here the results of performing forward and backward sweeps of
the grow rate $\mu$ in the quasi-one-dimensional situation described by
Eqs.\,\eqref{eq:1drho} and \eqref{eq:1dPx}. We do this following three
vertical lines in the phase diagram of Fig.\,\ref{fig:phase_diagram1d},
namely at $\mathrm{Pe}=0$, $\mathrm{Pe}=5$, and $\mathrm{Pe}=7.5$. The other parameters are kept fixed at
$\overline{D_r}=0.7$ and $L=10$. For each forward sweep, the system is
started into the homogeneous state $\rho=1$ and $p_x=0$, perturbed by
noise, and the equations are evolved in time until the transient is
gone, and the value of the spatially averaged density
$\langle\rho\rangle$ is recorded and plotted in Fig.
\ref{fig:histeresis}. Then, the value of $\mu$ is slightly increased,
the new transient allowed to vanish, and the new value of
$\langle\rho\rangle$ plotted again. This continuation procedure is
repeated until a large value of $\mu$, so that different instability
lines are crossed. Then, in the backward sweep the value of $\mu$ is
reduced in small steps, always using as initial condition the last state
for the previous value of $\mu$, and the corresponding values of
$\langle\rho\rangle$ plotted again.

We start with the case without self-propulsion, Pe=0. As seen in Fig.
\ref{fig:histeresis} (red triangles), for $\mu$ below the crossing with
the Turing instability line (indicated by the vertical solid line,
$\mu_c^{turing}$) the solution relaxes at long times to the stable
homogeneous solution ($\langle\rho\rangle=1$). Shortly after increasing
$\mu$ above $\mu_c^{turing}$,  $\langle\rho\rangle$ increases
continuously, indicating the development of spatially periodic patterns.
In the backward sweep, i.e. slowly decreasing $\mu$ from the previous
state, the system retraces precisely the states previously found.

For Pe=5 the behavior is different (large violet circles in Fig.
\ref{fig:histeresis}). As before, when increasing $\mu$ below the
predicted Turing instability line, which is indicated by the vertical
dotted line at  $\mu=\mu_c^{5.0}$, the system returns to the stable
homogeneous state characterized by $\langle\rho\rangle=1$. Shortly after
increasing $\mu$ above the Turing instability $\mu_c^{5.0}$,
$\langle\rho\rangle$ increases, but it does so undergoing the complex
transient depicted in Figs. \ref{fig:phases1d}c and \ref{fig:phases1d}g.
The final state is a travelling wave, with global polarization and a
relatively large value of $\langle\rho\rangle$. The result is a
discontinuous jump from $\langle\rho\rangle=1$ to this larger value when
crossing the Turing line. In Fig. \ref{fig:histeresis} this new state is
followed until large values of $\mu$, and then $\mu$ is decreased in a
backward sweep. When $\mu$ crosses again the $\mu=\mu_c^{5.0}$
transition, but this time from above, no qualitative change occurs: the
travelling wave solution exists and can be continued with this procedure
until a much smaller value of $\mu$. This indicates bistability between
the traveling wave and the homogeneous state in a range of values of
$\mu$ below the Turing line of Fig. \ref{fig:phase_diagram1d}, leading
to the corresponding hysteresis loop between the two attractors.

For Pe=7.5  (small yellow circles in Fig. \ref{fig:histeresis}) the
Turing line is not crossed, but rather the Hopf line. It is crossed at
$\mu=\mu_c^{hopf}$, a value marked by the vertical dashed line in Fig.
\ref{fig:histeresis}). By increasing $\mu$ below that line, the
homogeneous-density value $\langle\rho\rangle=1$ is recovered. After
increasing $\mu$ above the Hopf instability value $\mu_c^{hopf}$,
$\langle\rho\rangle$ increases discontinuously, probably indicating a
subcritical character for this Hopf bifurcation to travelling waves. By
continuing the forward sweep, and then reducing $\mu$ in the backward
sweep, we see that the state reached follows very closely the one
obtained for Pe=5 (violet circles), without qualitative changes when
crossing from above the Hopf line. This hysteresis indicates again
bistability between this wave state and the homogeneous one for a range
of values of $\mu$ below the Hopf line, and suggests that the
traveling-wave state found after a transient close to the Turing
instability arise from nonlinear interactions of the Turing and the Hopf
modes. The coexistence of attractors revealed in this Appendix is
probably just a small glimpse into a more complex solution structure
present in the one-dimensional and two-dimensional cases.

\begin{figure}[hbt!]
	\includegraphics[width=\columnwidth]{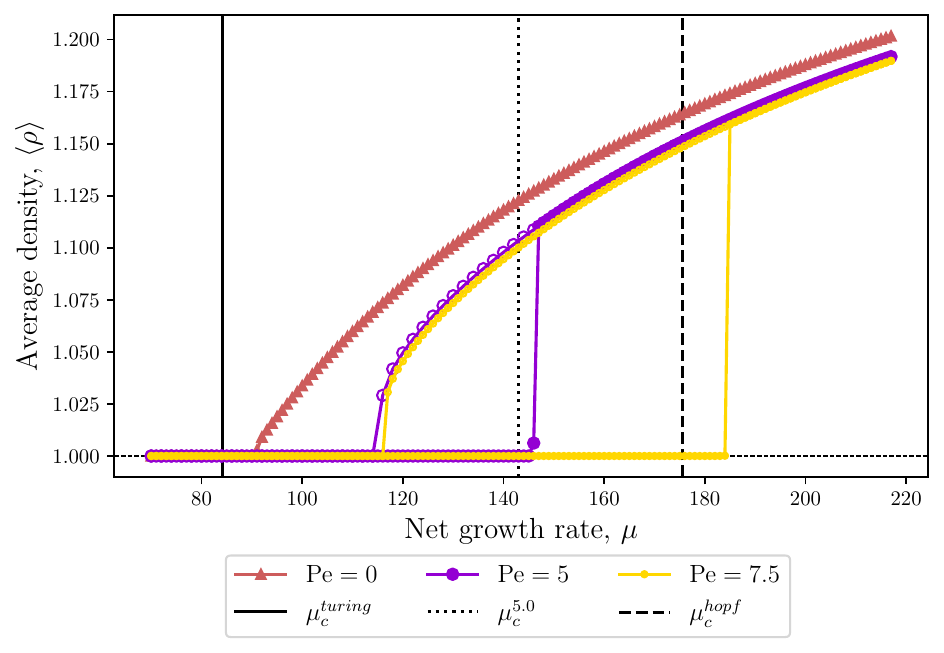}
	\caption{
		Spatially averaged density $\langle \rho\rangle$ of the
		quasi-one-dimensional system simulated from Eqs.\,\eqref{eq:1drho} and
		\eqref{eq:1dPx}. $L=10$ and $\overline{D_r}=0.7$. Whereas for Pe=0 (red
		triangles) a backwards sweep of the parameter $\mu$ retraces precisely
		the states found in a forward sweep, hysteretic behavior is found for
		Pe=5 (large violet circles) and Pe=7.5 (small yellow circles). This
		reveals bistability between traveling waves and homogeneous states. More
		details in the text.
	}
	\label{fig:histeresis}
\end{figure}

	\bibliography{Flocking_PRRv2}

\end{document}